\documentclass[11pt,a4paper]{article}

\usepackage{enumerate}
\usepackage{amssymb}
\usepackage{amsmath}
\usepackage{graphicx}
\usepackage{psfrag}
\usepackage[super]{natbib}

\newcommand{\Si}{$\Sigma$}
\newcommand{\shrteqn}[1]{
\begin{equation}
#1
\end{equation}
}

\newcommand{\eqnref}[1]{Eq. (\ref{#1})}
\newcommand{\eqnrefs}[1]{Eqs. (\ref{#1})}
\newcommand{\un}[1]{\underline{#1}}
\newcommand{\figref}[1]{Fig. \ref{#1}}
\newcommand{\dd}[2]{\frac{\partial #1}{\partial #2}}
\newcommand{\ic}{\frac{1}{c}}
\newcommand{\pr}[1]{${\rm #1}'$}
\newcommand{\fb}[1]{\boldsymbol {#1}}

\newcommand{\myfiga}{%
\begin{figure}
\begin{center}
\psfrag{theta1}{$\theta_1$}
\psfrag{Velocityv}{Velocity $v$}
\psfrag{t+delta1dash+delta1doubledash}{($t+\Delta'_1+\Delta''_1$)}
\psfrag{t+delta1dash}{$(t+\Delta'_1)$}
\psfrag{sigma1spath}{$\sigma_1$'s path}
\psfrag{sigma0spath}{$\sigma_0$'s path}
\psfrag{ptp}{$(t)$}
\includegraphics{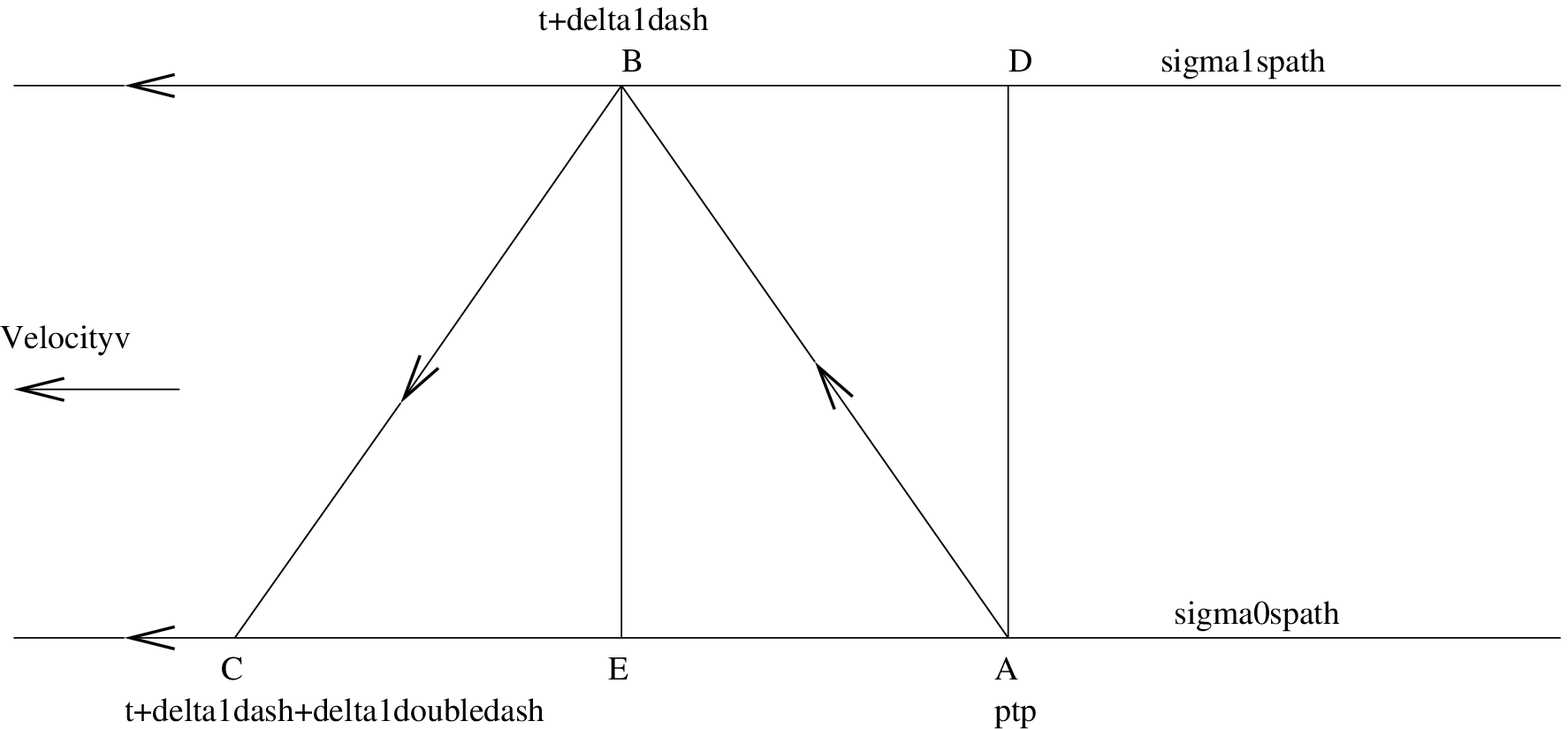}
\caption{Signal from $\sigma_0$ to $\sigma_1$ and back.}
\label{fig:fig1}
\end{center}
\end{figure}
}
\newcommand{\myfigb}{%
\begin{figure}
\begin{center}
\psfrag{t}{$t$}
\psfrag{t1}{ $t_1$}
\psfrag{t0}{ $t_0$}
\psfrag{tdashzero}{$t'_0$}
\psfrag{Pofpcommat}{P$(\fb{p},t)$}
\psfrag{sigma1colont1-delta1doubledash}{$\sigma_1$: $t_1-\Delta''_1$}
\psfrag{sigma0spath}{$\sigma_0$'s path}
\psfrag{Velocityv}{Velocity $v$}
\includegraphics{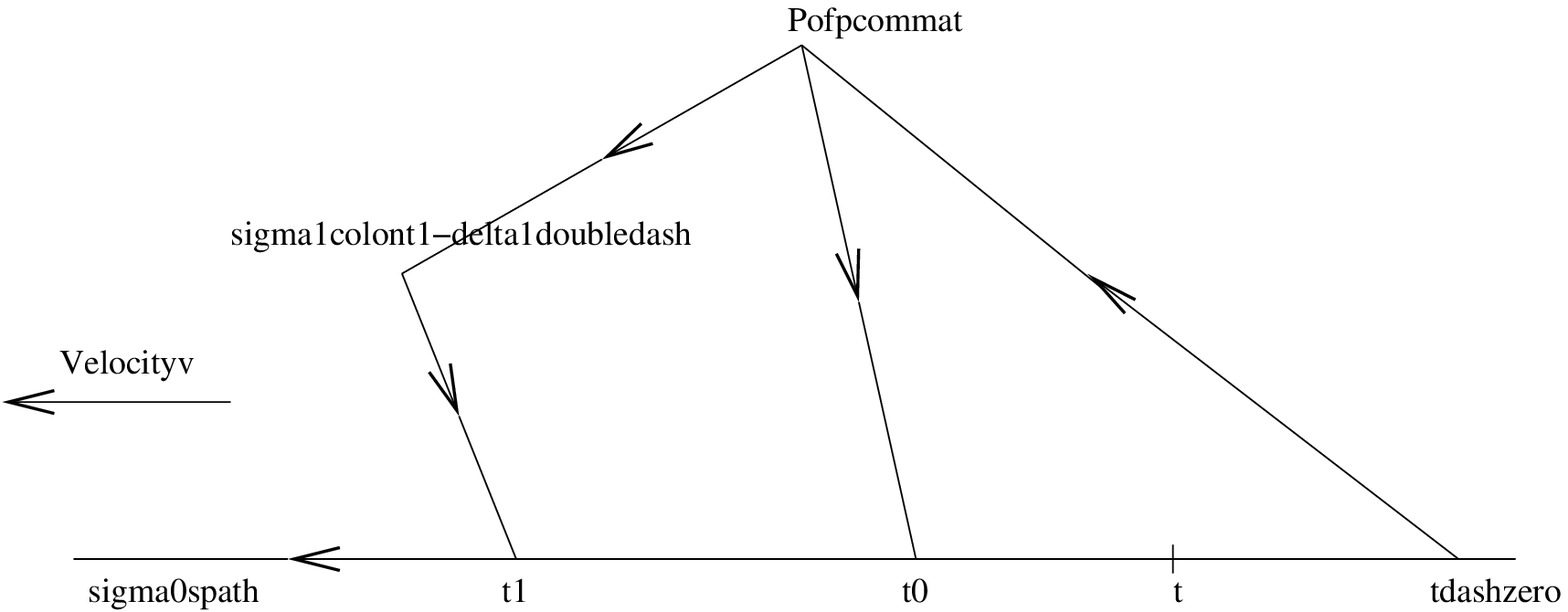}
\caption{Direct and indirect echo from P in S.}
\label{fig:fig2}
\end{center}
\end{figure}
}
\newcommand{\myfigc}{%
\begin{figure}
\begin{center}
\psfrag{picommatau}{P$(\fb{\pi}, \tau)$}
\psfrag{sigma0colontau0}{$\sigma_0$: $\beta_1 t_0$}
\psfrag{sigma0colontau1}{$\sigma_0$: $\beta_1 t_1$}
\psfrag{sigma1colontau1-delta1}{$\sigma_1$: $\beta_1 t_1-\delta_1$}
\includegraphics{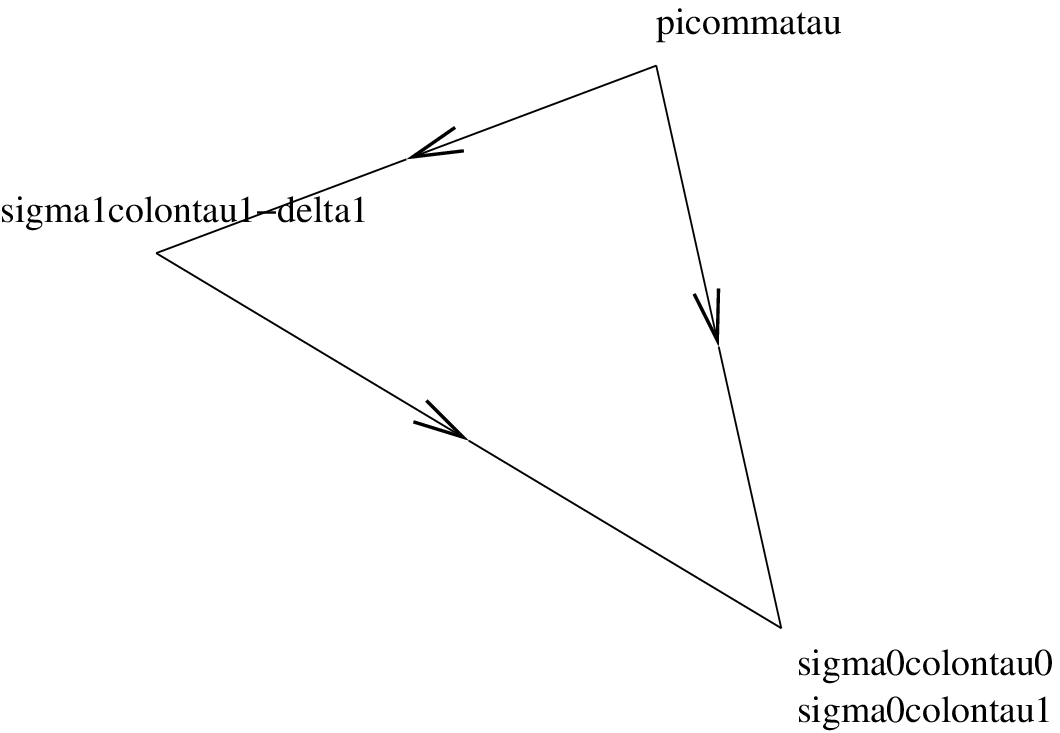}
\caption{Direct and indirect echo from P in $\Sigma$.}
\label{fig:fig3}
\end{center}
\end{figure}
}
\newcommand{\myfigd}{%
\begin{figure}
\begin{center}
\psfrag{theta1}{$\theta_1$}
\psfrag{theta2}{$\theta_2$}
\psfrag{theta12}{$\theta_{12}$}
\psfrag{v}{$v$}
\psfrag{sigma0spath}{$\sigma_0$'s path}
\psfrag{sigma1spath}{$\sigma_1$'s path}
\psfrag{sigma2spath}{$\sigma_2$'s path}
\includegraphics{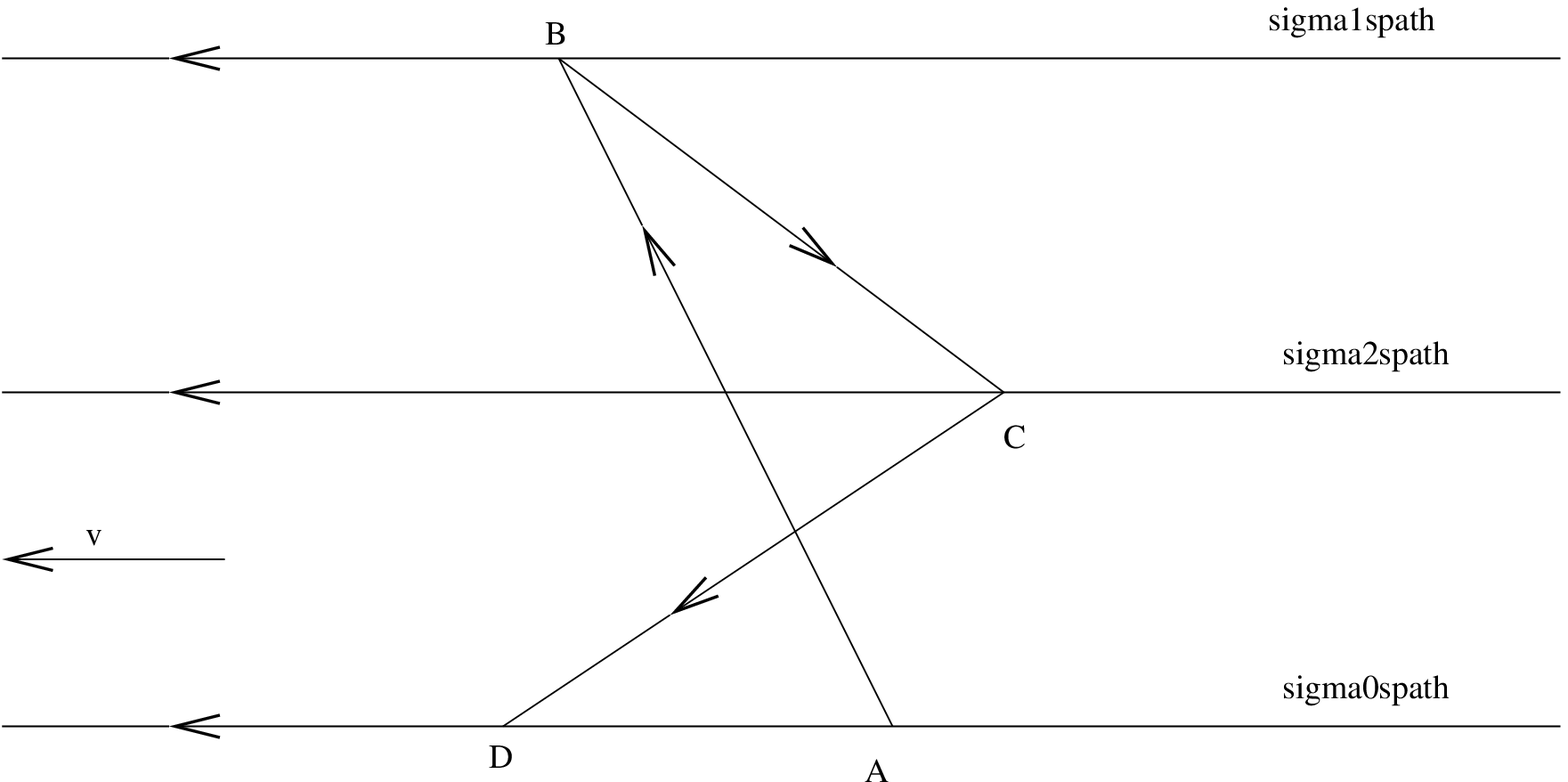}
\caption{Delays in a round trip in S.}
\label{fig:fig4}
\end{center}
\end{figure}
}

\begin{titlepage}
\title{EINSTEIN'S \ \ ``ZUR ELEKTRODYNAMIK...'' (1905) \ \ REVISITED, \ \ WITH \ \ SOME \ \  CONSEQUENCES}
\author{S. D. Agashe\\Adjunct Professor\\Department of Electrical Engineering\\Indian Institute of Technology\\Mumbai\\India - 400076\\email: eesdaia@ee.iitb.ac.in}
\date{}
\end{titlepage}

\begin{document}
\maketitle

\ 
\\
\\
Einstein, in his ``Zur Elektrodynamik bewegter K\"orper'', gave a physical (operational) meaning to  ``time'' of a remote event in describing ``motion'' by introducing the concept of  ``synchronous stationary clocks located at different places''. But with regard to ``place'' in describing motion, he assumed without analysis the concept of  a system of co-ordinates.
  
In the present paper, we propose a way of giving physical (operational) meaning to the concepts of  ``place'' and  ``co-ordinate system'', and show how the observer can define both the place and time of a remote event. Following Einstein, we consider another system ``in uniform motion of translation relatively to the former''.  Without assuming ``the properties of homogeneity which we attribute to space and time'', we show that the definitions of space and time in the two systems are linearly related. We deduce some novel consequences of our approach regarding faster-than-light observers and particles, ``one-way'' and ``two-way'' velocities of light, symmetry, the ``group property'' of inertial reference frames, length contraction and time dilatation, and the ``twin paradox''. Finally, we point out a flaw in Einstein's argument in the ``Electrodynamical Part'' of his paper and show that the Lorentz force formula and Einstein's formula for transformation of field quantities are mutually consistent. We show that for faster-than-light bodies, a simple modification of Planck's formula for mass suffices. (Except for the reference to Planck's formula, we restrict ourselves to Physics of 1905.)
\\
\\
Keywords: operational meaning, co-ordinate system, representation.

\section{EINSTEIN, RADAR AND GPS}
\subsection{Einstein's Synchronous Clocks}

 In the ``Kinematical Part'' of Einstein's celebrated ``Zur Elektrodynamik bewegter K\"orper''(all Einstein quotations are from the English translation in \cite{PRI}, except for two quotations from \cite{EINREL}), he remarked: ``If we wish to describe the \emph{motion} of a material point, we give the values of its co-ordinates as functions of the time. Now we must bear carefully in mind that a mathematical description of this kind has no physical meaning unless we are quite clear as to what we understand by ``time''''. He then introduced, ``with the help of certain imaginary physical experiments'', the concept of ``synchronous stationary clocks located at different places'', to enable one to determine the ``time'' of a remote event. This requires, however, setting up an infinitude of clocks located all over ``space'' and presumably also an infinitude of observers to read these clocks when events occur at their places. 

\subsection{Einstein's ``Co-ordinate System'' for Space}
	 Einstein assumed the availability of a ``system of coordinates'' with ``the employment of rigid standards of measurement and the methods of Euclidean geometry''. The most fundamental of these involve ``reaching out'' a remote place from an ``origin of co-ordinates''.  The suggestion that we might set up a ``a three-dimensional scaffolding of rigid meter sticks, with clocks for determining the time of local events situated at the nodal points'' \cite{BRIDGMAN}, or, more picturesquely,  ``Think of constructing a frame by assembling meter sticks into a cubical latticework similar to the ``jungle gym'' seen on playgrounds. At every intersection of this latticework fix a clock'' \cite{TAYLOR}, requires an infinitude of observers all over space to record the position of an event. Of course, the same observers could also read the clocks, thereby determining the space co-ordinates and time of occurrence of a remote event. (In a pre-Einsteinian method of place determination using the latticework, one would see a particle passing by a particular point on the latticework at a particular time in his watch, and then later on, go and check the co-ordinates of that point.) In his \cite{EINREL}, Einstein says: (p.6)``If, for instance, a cloud is hovering over Trafalgar Square, then we can determine its position relative to the surface of the earth by erecting a pole perpendicularly on the Square, so that it reaches the cloud. '' \ldots (p.7) ``We speak of the height of the cloud even when the pole which reaches the cloud has not been erected. By means of optical observations of the cloud from different positions on the ground, and taking into account the properties of the propagation of light, we determine the length of the pole we should have required in order to reach the cloud.''

\subsection{ Einstein's Approach and the ``Radar Approach'': Many Clocks or One Clock? }
Einstein's method of setting up synchronous clocks appears to be very similar to the so-called ``radar'' approach. Inspired by the acoustic phenomenon of an ``echo'', both involve the sending and receiving of a ``signal''. However, Einstein's \un{purpose} in setting up synchronous clocks was to provide a physically meaningful \un{definition} of the time of occurrence of a remote event. The radar approach, on the other hand was first used for the \un{detection} of a remote object, and later, for \un{ranging} - hence, \un{ra}dio \un{d}etection \un{a}nd \un{r}anging. Interestingly, the use of ``Hertzian waves'' for \un{ranging} was already envisaged by Nikola Tesla \cite{TES} (pp. 208-209) in 1900 : ``Stationary waves in the earth mean something more than mere telegraphy without wires to any distance. ... For instance, by their use we may produce at will,
from a sending-station, an electrical effect in any particular region of the globe; we may determine the relative position or course of a moving object, such as a vessel at sea, the distance traversed by the same, or its speed \ldots''. Soon thereafter, on April 30, 1904, a patent for the use of ``Hertzian'' waves for \un{detection} of a remote object was granted to Christian H\"ulsmeyer. (Of course, using radar, one can also determine ``time at a distance'', although, in practical applications, the distance is more crucial since the time is virtually that of the observation.) Unlike Einstein's approach, the radar approach requires only a \un{single} observer using only a \un{single} clock. Perhaps, Einstein was not aware of these developments. Or, if he was, he did not consider them in his study. In fact, Einstein did not use the idea of signaling as a means of determining the time of occurrence of a remote event since he assumed that the network of synchronous clocks had already been set up all over space. (In 1905, transmission and reception of electromagnetic waves was not a state-of-the-art task. In 2005, it is a commonplace.) Of course, setting up a synchronous clock or checking the synchronicity of clocks does require an echo.

	 Would it not be possible, using radar, not only to determine the distance and time of a remote event, but also to set up a co-ordinate system? If a directional antenna is used, the direction towards the remote object (line of sight) can also be determined by setting up three ``fixed'' ``reference'' directions or lines of sight, measuring angles, calculating direction cosines, etc. (In fact, it was this requirement that gave rise to the ``three reflecting stations'' idea of the present paper.) However, \un{books} on the Special Theory of Relativity which mention the radar approach have not spelt out the operational details of determination of place of the event, except in the case of one-dimensional motion in the context of another (moving) observer (for example, the ``k-calculus'' of Bondi\cite{BONDI}). Also, when the observing radar is mounted on a ``moving platform'', such as an aircraft, a ship or a land vehicle, the calculations are usually done without making relativistic correction. 

In the present paper, we show how, using only a single clock, a single ``stationary'' observer could \un{define} a co-ordinate system. What we mean is that a set of appropriate time observations made with a single clock may be related to or represented by points in 3-dimensional Euclidean geometry, a concept acquired through ``rigid standards of measurements'', or by triples of real numbers, a more abstract concept. In this sense we could talk of completing Einstein's kinematics. 
	 
\subsection{The GPS Approach for Time and Co-ordinate Determination }

There is also the \un{practical} method of location used in ``Global Positioning System''(GPS)\cite{LOGSDON}. It is based on the use of spatially separated synchronized clocks located in GPS SVs (Space Vehicles), a la Einstein. But peculiarly enough, in its use, instead of an observer determining the position and time of a remote event, we have an observer (GPS receiver) determining one's own location relative to some specific locations (the master control station and the monitor stations). It is not clear how a GPS could be used by a terrestrial observer to determine the position of a terrestrial or non-terrestrial event. GPS does not involve any echo measurements between SVs and the receiver. Our approach does have some similarity with the radar and GPS approaches, but we use it to \un{define} a co-ordinate system and also to study, like Einstein, the relationship between observations by two observers. The GPS method \un{presumes} a co-ordinate system but does make correction for relativistic effects (time dilatation and gravity). 

\subsection{Can We Do without an Echo?}
	  If the single observer could be ``assisted'' by three ``reflecting stations'' or ``repeater stations'', then by measuring the \un{time-differences} between the various signals, direct and indirect, from a remote object, could the observer determine the place and time of the object? We show below that it is almost possible to do so - almost, because in theory there are situations where a bivalent ambiguity may arise, i.e., two different determinations are possible for given data. (We give an example of this in Sec. 3.4.3) One would think that an additional reflecting station could be used to resolve the ambiguity, but it turns out that not even finitely many additional reflecting  stations would be able to resolve all possible ambiguities. However, if we allow the single observer to obtain an echo from a remote event, then with the help of the three reflecting stations, the place and time of the remote event can be uniquely determined. (An echo involves \un{sending} and \un{receiving} of a ``signal''. Note that for ``non-luminous'' objects, we would anyway need the echo approach, so well used by bats. Thus a signal sent by the observer to the non-luminous object is reflected back directly to the observer by the remote object, but it is also reflected by the latter to the three stations which, in turn, reflect it to the observer.) We emphasize that the time recording needs to be done only at one place, thus requiring only one clock, instead of an infinitude of synchronous clocks. The other stations serve merely to reflect the signal sent by the remote object. 
		
\subsection{What Is an ``Observer''?}
It would be correct to say that Einstein was the first to take seriously the concept of an ``observer''. Before his 1905 paper, there was, of course, talk of two ``co-ordinate systems'' or ``reference frames''. But even when talking about the ``Galilean transformation of co-ordinates'' given by $x'(t)=x(t)-vt$, $y'(t)=y(t)$, $z'(t)=z(t)$, $t'=t$, there was no explicit operational  characterization of the two observers involved therein. There was only one omniscient and omnipresent observer, looking at diagrams on paper !

\section{EINSTEIN'S TWO OBSERVERS}
\subsection{What is ``Place'' ?}
		In his paper, at the outset, Einstein emphasized that the notion of ``time'' in describing ``motion'' was not quite clear and needed a definition.  To this, we would like to add that even the notion of ``place'' is not clear and needs a definition.  Since this is a matter of definition, in our proposed definition, the question as to whether the set of stations, one of which will serve as an observer, are at rest or are moving together does not arise. (Synge \cite{SYNGE} goes so far as to say: ``Suppose that the event is the explosion of a rocket in mid-air. Let there be four observers, flying about in aeroplanes, not on any particular courses, but turning and diving and climbing in an arbitrary way. Let each observer carry a clock, not necessarily an accurate clock but perhaps an old battered clock - the one essential is that it keeps going. Each observer notes the reading of his clock when he hears the explosion of the rocket. Let these four readings be denoted by $(x^1,x^2,x^3,x^4)$; these four numbers may be taken as the coordinates of the event''.)  Of course, we do specify some \un{observable} requirements of the set of stations. Further, following Einstein, we do consider also the possibility of \un{another} set of stations, and thus, of defining a different ``time and space'' for remote events in exactly the same way as the first one. Naturally, the results of the observations by the two systems cannot be related unless one finds \un{out} or \un{assumes} exactly how the two systems themselves are related.  Thus, like Einstein, one may assume that the second system is moving uniformly relative to the first one.  In addition, one needs to make some assumptions about the behaviour of the signals themselves. One of these is what Einstein calls a ``postulate'', namely, that ``light is always propagated in empty space with a definite velocity c which is independent of the state of motion of the emitting body''.  His other postulate, which he called the ``Principle of Relativity'', is that ``to the first order of small quantities, the same laws of electrodynamics and optics will be valid for all frames of reference for which the equations of mechanics hold good''. 
(But, in a later section of \cite{PRI} Einstein states the Principle of Relativity as follows. ``The laws by which the states of physical systems undergo change are not affected, whether these changes of state be referred to the one or the other of two systems of co-ordinates in uniform translatory motion''.)
 Since we are looking at only \un{kinematical} considerations, we assume that the second system of observation and calculation be only similar to the first one in its operational aspects and do not consider laws of mechanics, leave alone laws of electrodynamics and optics, except towards the end of the present paper (Sec. 6.9).  Einstein assumed the notion of a co-ordinate system for space and so said: ``Let us in ``stationary'' space take two systems of co-ordinates, i.e., two systems, each of three rigid material lines, perpendicular to one another, and issuing from a point'' without operationally specifying the meaning of ``rigid''. When talking about ``another system in uniform motion of translation relatively to the former'' he said: ``Now to the origin of one of the two systems (k) let a constant velocity $v$ be imparted in the direction of the increasing $x$ of the other stationary system (K), and let this velocity be communicated to the axes of the co-ordinates, the relevant measuring-rod, and the clocks''.  In our approach, we simply assume that we have one	 system somehow given or set up, and that we have another system somehow set up which the observer of the first system finds out, on the basis of observations, to be in uniform motion - ``relative'' to it, of course.
(If the other observer is another radar, then that has to have its own set of reference directions. We have not seen any discussion, involving relativistic considerations, of the \un{same} object being sighted simultaneously by two radars, not stationary relative to one another.)

\subsection{What is ``the same Event''?}
 Of course, as Bridgman has remarked, the concept of the ``same event'' being observed by two systems of observation is not operationally clear, and lurking behind it may be the idea of ``absolute time and space''.  Thus, one talks about a ``lightning flash'' as an event, but how do the two systems of observers \un{know} that they are observing the same event to which they assign perhaps different times and places?  Perhaps, a ``collision of two particles'' or the ``onset of a lunar eclipse'' are the sort of event about which two observers may agree that they are observing the ``same'' event.

 Einstein tacitly assumes that there must be some definite relations between the findings of the two observers : ``To any system of values $x$, $y$, $z$, $t$, which completely defines the place and time of an event in the stationary system, there belongs a system of values $\xi$, $\eta$, $\zeta$, $\tau$, determining that event relatively to the system $k$ \ldots''.  Whether or not one should \un{assume} the \un{same} constant ``velocity'' of the signal  (light) c in the different systems is perhaps a matter of choice, although Einstein \un{deduces} ``as required by the principle of the constancy of the velocity of light, in combination with the principle of relativity'' that ``light is also propagated with velocity c when measured in the moving system''. We feel that the velocity of light is not a Law of Physics; it could be considered to be a ``parameter''.

\section{SOME (PHILOSOPHICAL) CLARIFICATIONS}
\label{sec:Phil}
\subsection{Are the Considerations here Based on some ``Philosophy'' of Time? }
  We assume a minimum, commonsensical, philosophy of time, namely, that human beings have experiences of ``moment'' (``at'', ``when''), duration (``while'', ``during'') and tense (present, past, future), and that they have available some ``local'' method of observing the time-instant when they have some (momentary) experience. For example, we may assume an ``analogue'' clock-face or a digital display placed ``very close'' to the eyes of the observer. It could be the ``geodesic clock'' of Marzke \cite{MARZKE,OHANIAN,BASRI}. Of course, we do not suppose that anything will do under the name of a clock.  Certainly, the entire past history and experience of mankind has contributed to the concept of time and development of an artifact called a ``clock''.  Today, we have ``atomic clocks'' (as in GPS SVs) which are very ``stable'' and so will remain in synchronism over a very long duration. For those who are so inclined, we could say that a time-instant is represented by an element of the set R of real numbers and that to each momentary experience of an observer, a unique time-instant is assigned. 
In principle, one could allow a very crude ``clock''- think of Galileo counting his pulse beats, or, a child reciting the number-words ``one'', ``two'', ``three'', \ldots, when playing a game of hide-and-seek.

Interestingly, most authors, when they talk about a clock, think of some repetitive or periodic process. Synge \cite{SYNGE} says (p.14): ``To measure time one must use a \emph{clock}, a mechanism of some sort in which a certain process is repeated over and over again under the same conditions, as far as possible. The mechanism may be a pendulum, a balance wheel with a spring, an electric circuit, or some other oscillating system \ldots''. Such a clock is a discrete one and requires a counter. Instead, one could think of a capacitor being charged or discharged very slowly through a resistor, or better still, of the decay of radioactive carbon! Ideally, the process would go on forever.

\subsection{ Are the Considerations  Based on some ``Philosophy'' of Space ? }
 	Again, we assume a minimum, commonsensical, philosophy of space, namely, that human beings have experiences of place (``at'', ``here'', ``there'') and distance (``near'', ``far'') through the various senses such as sight, hearing, and touch, and also through their own mobility. We also realize that mankind has developed methods of determination of distance and position, such as by stretching a rope or a cord or the chain of a surveyor, using a ``rigid'' rod, sighting through a surveyor's ``level'' or a theodolite, measuring parallax, etc.  But in our approach we take the view that the basic observations to be made are only of the time of transmission and reception of ``signals'' by only one observer. We show that it may be possible to \un{represent} these by points in 3-dimensional Euclidean geometry, or, more abstractly, by triples of real numbers, i.e., by elements of the set $\mathbb{R}^3$, or, even more abstractly, by elements of an inner product space. This representation, possible in infinitely many different ways, may be called ``co-ordinatization'' of the events, or ``setting up a co-ordinate system''. To repeat, we assume that what is \un{observed and recorded} is an experience of time; what is \un{defined and calculated} is a representation in a manner which is certainly influenced by our experience of space.
   We are certainly not entertaining any conception of a ``space-time continuum'' or of ``spacetime''.  Rather than consider time as a fourth dimension of space, we prefer to consider space - as far as our representation is concerned - as three additional dimensions of time!  We do have experiences of seeing remote ``objects'' such as a flying aircraft, the moon, and even galaxies, which cannot be reached by stretching a cord, or by laying out repeatedly a rigid rod, or by moving out to the object.  Yet we seem to want to extend our concepts of place and distance, based on stretched cords, rigid rods and moving from one place to another, to these remote objects. We show how this can be done.

Interestingly, in \cite{EINREL} (p.9) Einstein wrote: ``In the first place we entirely shun the vague word ``space'', of which, we must honestly acknowledge, we cannot form the slightest conception, and we replace it by ``motion relative to a practically rigid body of reference'' ''. However, in \cite{PRI}, when deriving the ``transformation of co-ordinates and times'', he appealed to ``the properties of homogeneity which we attribute to space and time''.

\subsection{ Are the Considerations Based on some ``Philosophy'' of Light (as a Signal)? }
 With Einstein, we do say that light ``travels'' from one ``place'' at one ``time'' to another ``place'' at another ``time'' with a constant ``velocity'' irrespective of what body emits the light and what observer, or system of observations, is used to receive the signal. Of course, in our view, what is \un{observed} is only the transmission and reception of the signal at the time shown by the clock.  The \un{other} times and places of the signal are only \un{inferred}, i.e., defined and calculated.  Thus, in our proposal, the observer receives \un{four} signals, emitted by an object - we are tempted to say, at some place and time -  one of them coming directly to the observer from the emitting object, and the other three, after reflection at three stations. Einstein calls them ``light signals'' or ``light stimuli'' \cite{EINREL}. We could think of them as flashes or pulses of light, or even as ``light particles''.

  Unlike Einstein, however, we do not assume that velocity of light has been ascertained by experiment to be such and such, because that will involve previous determination of both time and space.  Rather, we take velocity as a mere constant that enables us to define distance in terms of time, much like the astronomical way of using light-years.  Indeed, we could even take the ``velocity'' to be unity so that the distance traveled by a light ray is just another name for the duration of its travel. A distance for us is basically associated with the time of ``travel'' of light.

\subsection{Are the Considerations Based on some ``Philosophy'' of Geometry?}
\label{subsec:3d}
  Specifically, are we assuming some ``geometry'' of ``space''?  For example, are we assuming the geometry of space to be ``Euclidean'' and 3-dimensional?  What we shall use is a simple kind of ``distance'' or ``metric'' geometry \cite{MEN} wherein we have ``points'' and ``distances between them'' - which are non-negative real numbers - satisfying the usual ``metric space'' axioms, in particular, the ``triangle inequality''.   So it seems that the ``geometry'' of an \un{inner product space} is adequate. However, as we have emphasized, the choice of a ``geometry'' for space is only a choice of representation.

First, we need certain extensions to 3-dimensional or ``solid'' geometry of some Euclidean results in 2-dimensional or ``plane'' geometry. 

\subsubsection{Results from 2-dimensional or ``plane'' Geometry: Triangle Inequalities}

\un{Euclid's ``Elements'' I.20} states : ``In any triangle, two sides taken together in any manner are greater than the remaining one.'' (Hence the term ``triangle inequality''.)
 
Then, \un{I.22}: ``Out of three straight lines, which are equal to three given straight lines, to construct a triangle: thus it is necessary that two of the three (given) straight lines taken together in any manner should be greater than the remaining one.'' (It is enough to check that the longest of the three lines is less than the other two lines taken together.)

 To construct the triangle, Euclid has to draw or construct some circles. If the construction takes place in a plane, then there are \un{two} triangles that satisfy the requirement (with differing ``orientation''). What Euclid shows is that we can ``locate'' three points A, B, C, in a plane such that the lines joining them, AB, BC, CA, are ``equal'' to the three given lines. After joining the points (vertices), a triangle is obtained (with sides in addition to the vertices). If not all the triangles inequalities are satisfied, there may no such triangle, or the triangle may be a ``degenerate'' one, with A, B, C being collinear. 

We can show easily that any (non-degenerate) triangle can be co-ordinatized, i.e., \un{represented}, non-uniquely in $\mathbb{R}^2$. (We say $\mathbb{R}^2$ and not ``a plane'', because this can be done without assuming that the triangle ``lies'' in a plane.) If ABC is the triangle, represent A by (0,0), B by (AB, 0) and C by (x, y), with x, y chosen suitably, in two different ways. Of course, this is only one representation and assumes a definition of ``distance'' in $\mathbb{R}^2$. Such a co-ordinatization can be given a familiar visual meaning : choose in a ``plane'' the point A as the origin of co-ordinates, any line $\rm{X'AX}$ through A and B as the x-axis and a line $\rm{Y'AY}$ perpendicular to the x-axis as the y-axis. 

We point out the following abstract, metric-space counterpart of I.22 and representation in $\mathbb{R}^2$. If $\{\rm A,\rm B,\rm C\}$ is a set on which there is a metric $\rho$ then there is a representing function $\phi$: $\{\rm A,\rm B,\rm C\} \rightarrow \mathbb{R}^2$ such that $\rho(\rm A, \rm B) = d_e(\phi(\rm A), \phi(\rm B))$, where $d_e$ denotes the ``Euclidean'' distance in  $\mathbb{R}^2$. Briefly, a metric space with three elements can be ``embedded'' in $\mathbb{R}^2$. Note that it may be possible to embed it in $\mathbb{R}$, corresponding to the geometric situation when A, B, C are collinear. Instead of $\mathbb{R}^2$, we could use a two-dimensional inner-product space.

\subsubsection{Extension to 3-dimensional or ``solid'' Geometry: Tetrahedral Inequalities}

We need the extension of the above two Euclidean propositions to solid geometry, specifically, for a tetrahedron, i.e., a problem with six lines (and four points). Obviously, in any tetrahedron, the three sides of each face of the tetrahedron have to satisfy I.20. But what about the counterpart of I.22? It seems that there is a partial counterpart to this. (We have not seen this result stated as a theorem in \un{axiomatic} treatments of solid geometry.)

\un{Partial Counterpart}: Given six straight lines, suppose that some three of them satisfy I.22 and thus can be made into a triangle, say, ABC. If the remaining three straight lines satisfy appropriate inequalities, then a tetrahedron can be erected on triangle ABC such that these remaining straight lines are equal to the three edges of the tetrahedron other than those of the base ABC. Thus, denoting the would-be vertex by D, we have three additional triangles DAB, DBC, DCA, and so with some choice of the three remaining straight lines as the ``proposed'' edges DA, DB, DC, the appropriate triangle inequalities must be satisfied for the ``proposed'' faces DAB, DBC, DCA. (To construct the tetrahedron in space, i.e., to ``locate'' the fourth vertex, one would need to construct appropriate spheres and two different tetrahedra would result.)

Again, we can show easily that the vertices of any (non-degenerate) tetrahedron can be co-ordinatized, i.e., represented non-uniquely in $\mathbb{R}^3$ without visualizing $\mathbb{R}^3$ in terms of three co-ordinate axes . If ABCD is the tetrahedron, represent A by $(0,0,0)$, B by $(\rm{AB}, 0,0)$, C by $(x_1, y_1, 0)$ for suitable non-unique $x_1$, $y_1$, and D by $(x_2, y_2, z_2)$ for suitable non-unique $x_2$, $y_2$, $z_2$. Again, this is only one possible representation. There is, again a visual meaning that can be given to this co-ordinatization.

We have the following metric space counterpart : If $\{\rm A,\rm B,\rm C, \rm D\}$ is a set on which there is a metric $\rho$ then there is a representing function $\phi$: $\{\rm A,\rm B,\rm C, \rm D\} \rightarrow \mathbb{R}^3$ such that $\rho(\rm A, \rm B) = d_e(\phi(\rm A), \phi(\rm B))$, where $d_e$ denotes the ``Euclidean'' distance in  $\mathbb{R}^3$. Briefly, a metric space with four elements can be ``embedded'' in $\mathbb{R}^3$. Note that it may be possible to embed it in $\mathbb{R}^2$, corresponding to the geometric situation when A, B, C, D are coplanar. Instead of $\mathbb{R}^3$, we could use a three-dimensional inner-product space.

\subsubsection{Representability of additional Points}
The following problem of representability of additional points can arise in the plane, i.e., in $\mathbb{R}^2$. Suppose ABC is a given (non-degenerate) triangle (i.e., suppose three straight lines satisfying the triangle inequality are given), and we have a representation of it in $\mathbb{R}^2$. Suppose a fourth point D is given, or, rather three more straight lines DA, DB, DC are given, such that the triangle inequalities are satisfied for the triangles DAB, DBC, DCA. Is D representable in $\mathbb{R}^2$? (Equivalently, is D coplanar with ABC?) The answer is, of course, that D is not necessarily representable in $\mathbb{R}^2$, since a point D can be chosen which is \un{not} coplanar with A, B, C. (This is an axiom of ``solid'' geometry.) If D is representable, its representation is unique. The metric space counterpart of this is that a four-element metric space may not be embeddable in $\mathbb{R}^2$.

Now consider the counterpart of this in solid geometry. Suppose a (non-degenerate) tetrahedron DABC is given and which is, therefore, representable in $\mathbb{R}^3$. Suppose a fifth point E is given and four more straight lines are given which are to be the sides EA, EB, EC, ED. Is the point E co-ordinatizable, i.e., representable in $\mathbb{R}^3$? Obviously, these new straight lines must satisfy the triangle inequalities for the new triangles that are to be formed. But are these inequalities sufficient to guarantee the representability of E as a point of 3-dimensional space?  If not, one would have to say that the point E is not representable in $\mathbb{R}^3$, i.e., in 3-dimensional space, and so, one could look for representability in  $\mathbb{R}^4$. Thus, a five-element metric space of which a four element subset is embeddable in $\mathbb{R}^3$ may not be embeddable in $\mathbb{R}^3$. 

To construct a counterexample for  representability in  $\mathbb{R}^3$, we go to a fourth dimension and choose five appropriate 4-tuples in $\mathbb{R}^4$ such that four of them form a tetrahedron, and so, this tetrahedron can be represented in $\mathbb{R}^3$. But the fifth ``point'' cannot be represented in $\mathbb{R}^3$. (Counterexample 1: choose the 4-tuples as follows. A:$(0,0,0,0)$, B:$(1,0,0,0)$, C:$(0,1,0,0)$, D:$(0,0,1,0)$, E:$(0,0,0,a)$, with $a\ne 0$ . The various distances are: $\rm{AB}=\rm{AC}=\rm{AD}=1$, $\rm{AE}=a$, $\rm{BC}=\rm{BD}=\rm{CD}=\sqrt{2}$, $\rm{BE}=\rm{CE}=\rm{DE}=\sqrt{a^2+1}$. The triangle inequalities are satisfied for all the triangles, namely, ABC, ABD, etc. Now, A, B, C, D can obviously be represented in $\mathbb{R}^3$ as $(0,0,0), (1,0,0), (0,1,0), (0,0,1)$ respectively, but with this representation we show that E cannot be represented in $\mathbb{R}^3$. We have $\rm{BE}^2=\rm{AB}^2+\rm{AE}^2$, so AB is perpendicular to AE in $\mathbb{R}^3$, as are also AC, AD, which is not possible in $\mathbb{R}^3$.) If, however, instead of four straight lines or lengths EA, EB, EC, ED, three differences in lengths, say, EB - EA, EC - EA, ED - EA are specified such that these satisfy appropriate triangle inequalities, then the problem has a (non-unique) solution. (We give an example of non-uniqueness. Counterexample 2: let A be $(0,0,0)$, B:$(1,0,0)$, C:$(0,1,0)$, D:$(0,0,1)$, E:$(-0.1702,-0.1702,-0.1702)$, $\rm{E}'$:$(0.0373, 0.0373, 0.0373)$. Then $\rm{EA}=0.2948$, $\rm{E'A}=0.0646$, so $\rm{EA} \ne \rm{E'A}$ but $\rm{EB}-\rm{EA} = \rm{E'B}-\rm{E'A} = \rm{EC}-\rm{EA} = \rm{E'C}-\rm{E'A} = \rm{ED}-\rm{EA} = \rm{E'D}-\rm{E'A} = 0.9$.) 

We note that in our approach the co-ordinatization or representation is not any ``intrinsic'' property of ``space'' and we are not assuming that space ``has'' a particular ``metric''. We are simply choosing a representation which is convenient (and familiar)!

\section{THE NEW APPROACH}
In the new approach proposed here for the \un{definition and calculation} of \un{both} time and space co-ordinates, we assume a system S consisting of one observer $s_0$ with a clock and three reflecting ``stations'' $s_1$, $s_2$, $s_3$. (Einstein used the letter K to denote what he called a ``stationary'' system and letter $k$ to denote another ``moving'' system. We will use the corresponding Greek letter $\Sigma$ to denote the other observation system.) Suppose the observer $s_0$ observes four time-instants in his clock: one, $t_0$, of direct reception of a signal emitted by a distant object P when something happens; another,a time instant $t_1$ of reception of a signal via, i.e., after reflection at, $s_1$, and similarly, instants $t_2$, $t_3$. (Thus, this may correspond to $s_0$ ``seeing'' a flash ``directly'' at time $t_0$, and then seeing images of the ``same'' flash in the ``mirrors'' at $s_1$, $s_2$, $s_3$ at instants $t_1$, $t_2$, $t_3$.) How shall we \un{define} the place and time of occurrence of this event?

It cannot be overemphasized that we are trying to \un{propose} a \un{definition} of the space co-ordinates and time of an arbitrary event on the basis of observed time instants $t_0$, $t_1$, $t_2$, $t_3$. Of course, we do not want to do this arbitrarily. (With complete arbitrariness, as mentioned by Synge \cite{SYNGE}, there may not be much that we can say.) In particular, we do hope that the proposed definition will correspond, when feasible, to the classical definition achieved with ``the employment of rigid standards of measurements and the methods of Euclidean geometry''. So, we look for \un{a} co-ordinatization or \un{representation} in $\mathbb{R}^3$ (and a little more generally, in a three dimensional inner-product space, say L, over $\mathbb{R}$) of the ``space'' aspect of an event, and in $\mathbb{R}$ of the ``time'' aspect of the event. Further, we do not want the reflecting stations $s_1$, $s_2$, $s_3$ to ``behave'' in any arbitrary manner. We, therefore, assume that the four stations $s_0$, $s_1$, $s_2$, $s_3$ form a ``rigid'' system as evidenced observationally. Thus, we assume that by using the ``echo'' method, the observer $s_0$ ascertains that the three stations are at a constant delay from $s_0$ and from one another - recall that for us, a distance is a time-difference. We assume that the delays between the reflecting stations are symmetric, i.e., the delay from $s_1$ to $s_2$, say, is equal to the delay from $s_2$ to $s_1$, and so on. Let these one-way time delays between $s_0$ and $s_1$, $s_2$, $s_3$ be denoted by $d_1$, $d_2$, $d_3$, and the time delay between $s_1$ and $s_2$, ascertained indirectly, by $d_{12}$, etc. We assume that the stations form a non-degenerate tetrahedron, and that the appropriate triangle inequalities are satisfied. (Of course, this can be verified knowing the distances $d_1$, etc., and $d_{12}$, etc., and we expect this to happen because of our beliefs that light takes the shortest path between two points, and that the shortest path between two points is a straight line.) 

The observer now postulates that the signal was emitted by the remote object at some (unknown) ``time'' $t$, and thus traveled from the object at time $t$ to the observer at time $t_0$, with $t_0\ge t$, and thus the ``distance'' between P and $s_0$ is $(t_0-t)$, choosing the velocity of light as ``1'' i.e., expressing distance in terms of ``light-time''.  Similarly, the signal sent by P at time $t$ must have reached the reflecting station $s_1$ at time $(t_1-d_1)$, with $(t_1-d_1) \ge t$, so that after reflection at $s_1$, it reached $s_0$ at time $(t_1-d_1)+d_1$, i.e., $t_1$. So the distance between P and $s_1$ is $(t_1-d_1-t)$. Likewise for the signals received from the other stations. Note that we have to honestly admit that the observer at $s_0$ cannot ``see'' the signal (light ray or light particle) leaving P and arriving at his own place; he imagines or assumes the signal to ``leave'' and ``arrive'', but, of course, he does see it at his own place. Similarly, $s_0$ does not see the signal leaving at a remote place P at one time, arriving at the remote reflecting station $s_1$, say, at another time, getting reflected instantaneously and arriving at his own place. So, what we are assuming is that if we imagine (assume) a light signal to leave ``place'' P at ''time'' $t_P$ and to arrive directly at ``place'' Q at ''time'' $t_Q$, then the ``distance'' between P and Q shall equal the duration between $t_P$ and $t_Q$ (``velocity'' of light = 1). At this stage, we are not assuming any specific ``path'' for the signal between P and Q. There are, of course, practical problems that may arise. One may not be able to receive a particular signal at all - one will then ascribe it to an ``obstruction''.

Now, knowing the ``distances'' $d_1$, $d_2$, $d_3$, and $d_{12}$, $d_{13}$, $d_{23}$, we can co-ordinatize, i.e., represent, these stations $s_0$, $s_1$, $s_2$, $s_3$ non-uniquely in $\mathbb{R}^3$. In particular we may represent $s_0$ by the origin $(0,0,0)$ of the co-ordinate system. One could also represent them by vectors in a 3-dimensional inner product space, say L, and thus, in particular, $s_0$ by the zero vector of the vector space. (In 1905, physicists were not very familiar with the concept of an \un{abstract} inner product space. Even mathematicians were only beginning to get to know it. However, physicists were familiar with the concept of n-dimensional Euclidean space.) We will use this abstract representation in our derivations below because, today, physicists are quite familar with the concept of an abstract inner product space. We will denote the representations of the stations $s_0$, etc., in L by the same symbols in boldface, $\fb{s_0}$, etc. The fact that the four stations form a non-degenerate tetrahedron implies that the three vectors $\fb {s_1}$, $\fb {s_2}$, $\fb{ s_3}$ form a basis for L, with $\fb{s_0} = \fb{0}$.

The problem then would be to determine the representation $\fb {p}$ in L of the space aspect of the event P (the co-ordinate triple $(x,y,z)$ if L is $\mathbb{R}^3$), and the unknown time $t$, from the measured time instants $t_0$, $t_1$, $t_2$, $t_3$. That is, we have to determine a vector $\fb {p}$, and a number $t$ (or, equally well, the number $(t_0-t)$ ) such that 
\shrteqn{
\begin{array}{lllll}
||\fb {p}-\fb {s_0}||& =& t_0 - t& \ge& 0,\\
||\fb {p}-\fb {s_i}||& =& t_i - d_i - t& \ge& 0,\ \ \ i=1,2,3, 
\end{array}
\label{eqn:a1}
}
where $||\cdot||$ denotes the norm of a vector. Hopefully, this problem has a solution (there are 4 unknowns and 4 equations) and a unique one. Note that $||\fb {s_1}|| = d_1$, etc., $||\fb {s_1}-\fb {s_2}|| = d_{12}$, etc. (A similar set of equations arises in GPS, except that usually one finds it stated that only three distances are enough. But this is because the GPS receiver is known to be on one particular side of the triangle formed by the three SVs.) Would the determined time $t$ and the norm $||\fb {p}||$ depend on the choice of the space L and on the embedding of the stations in L ? Indeed, we will show that they do not. (Recall that the station $s_0$ is represented by the zero vector of the vector space L.)

We assume that these time instants ${t_i}$ satisfy certain additional inequalities, other than the obvious ones given above, which follow from the triangle inequalities for the various triangles formed by P and the four stations. Thus, for example, for the triangle $\rm{P}s_0s_1$ we have $\rm{P}s_1 + s_1s_0 \ge \rm{P}s_0$ so that
\[
(t_1-d_1-t) + d_1 \ge (t_0-t)
\]
and so 
\[
t_1\ge t_0,
\]
which is what we expect since, the signal goes \un{directly} to $s_0$ at time $t_0$ and \un{indirectly} via $s_1$ at time $t_1$. 
We also have
\[
\rm{P}s_0 + s_0s_1 \ge \rm{P}s_1
\implies (t_0-t) + d_1 \ge t_1 - d_1 - t,
\]
and so
\[
t_1 - t_0 \le 2d_1.
\]
As remarked in Sec. \ref{subsec:3d} above, this problem may not have a solution -  unless we believe, with Einstein, that ``space'' \un{is} 3-dimensional. If it does not have a solution, either we could say that ``space'' is not 3-dimensional, or we could say that the paths of the light particles may not be straight lines - perhaps because of the effect of gravity (but why not of an electromagnetic field?). (In our approach, we could also think of setting up one more reflecting station. Only experience can show whether that will suffice to represent the observations we actually make! Our calculations below can be easily extended to handle more that three reflecting stations.) Assuming that it has a solution, we show that it will have two solutions, and we will have to choose the one which satisfies the inequalities such as $(t_0 - t) \ge 0$ and others above. 

To solve the \eqnrefs{eqn:a1}, squaring the equations we get (since $\fb{s_0}$ is the zero vector):
\shrteqn{
\begin{array}{lll}
||\fb {p}||^2& =& (t_0 - t)^2,\\
||\fb {p}- \fb {s_i}||^2& =& (t_i - d_i - t)^2, \ \ \ i=1,2,3.
\end{array}
\label{eqn:a1sqr}
}
We will denote the inner product of two vectors $\fb {v_1}$, $\fb {v_2}$ in L by $(\fb {v_1} \cdot \fb {v_2})$ or $\fb {v_1} \cdot \fb {v_2}$. Further, we will denote, for a vector $\fb {v}$ in L, $||\fb {v}||^2$ by $v^2$, or, occasionally, by $\bf{v}^2$.

Since $\{\fb {s_1}, \fb {s_2}, \fb {s_3}\}$ is a basis of L, we have a basis expansion for $\fb {p}$: 
\shrteqn{
\fb {p} = \alpha_1 \fb {s_1} + \alpha_2 \fb {s_2} + \alpha_3 \fb {s_3}.
\label{eqn:a2}
}
Denoting $(t_0-t)$ by $d$, we get :
\shrteqn{
\begin{array}{lll}
p^2&=&d^2,\\
p^2-2 \fb {s_i}\cdot \fb {p} + s_i^2 & = & (t_i - d_i - t_0 + d)^2, \ \ \ i=1,2,3.
\end{array}
}
By appropriate subtraction, we eliminate both $p^2$ and $d^2$ to obtain:
\shrteqn{
\begin{array}{lll}
\fb {s_i}\cdot \fb {p} & = & (t_i - d_i - t_0) d + \frac{1}{2} [s_i^2 - (t_i - d_i - t_0)^2], \ \ \ i=1,2,3.
\end{array}
\label{eqn:a3}
}
Let $G_s$ be the Gram matrix of the three vectors $\fb {s_1}$, $\fb {s_2}$, $\fb {s_3}$: 
\[
(G_s)_{ij}=\fb {s_i}\cdot \fb {s_j}.
\]
Note that since $\{\fb {s_1}, \fb {s_2}, \fb {s_3}\}$ is a basis for L, $G_s$ is positive definite. 

Since the vector $\fb {s_i}$ represents the station $s_i$ which is at a delay of $d_i$ from the station $s_0$, we have : 
\shrteqn{
\fb {s_i} \cdot \fb {s_i} = ||\fb {s_i}||^2 = d_i^2,
}
and since $s_i$ is at a delay $d_{ij}$ from $s_j$, we have : 
\shrteqn{
\begin{array}{lll}
\fb {s_i} \cdot \fb {s_j} & = & \frac{1}{2} [(||\fb {s_i} - \fb {s_j}||)^2 - s_i^2 - s_j^2],\\
&=&\frac{1}{2} (d_{ij}^2 - d_i^2 - d_j^2).
\end{array}
}
Hence, the entries of the matrix $G_s$ are independent of the choice of the representing vectors in L and, indeed, of the choice of the space L itself, but depend only on the delays $d_i$ and $d_{ij}$. This fact will enable us to show that if \eqnrefs{eqn:a1} have a solution, then $t$ and, therefore, $||\fb {p}||=t-t_0$, will not depend on the choice of the representation in L, nor on the choice of the space L.

Let $\alpha$ denote the column $[\alpha_1\ \ \alpha_2\ \ \alpha_3]^T$. Then, from \eqnrefs{eqn:a2} and \eqnref{eqn:a3}, we obtain : 
\shrteqn{
G_s \alpha = da+b
\label{eqn:a4}
}
where $a,b$ are the columns given by 
\shrteqn{
\begin{array}{lll}
a & = & [(t_1 - d_1 - t_0)\ \ \ (t_2 - d_2 - t_0)\ \ \ (t_3 - d_3 - t_0)]^T,\\
b & = & \frac{1}{2}[s_1^2-(t_1 - d_1 - t_0)^2\ \ \ s_2^2-(t_2 - d_2 - t_0)^2\ \ \ s_3^2(t_3 - d_3 - t_0)^2]^T.
\end{array}
}
Note that the columns $a$ and $b$ are independent of the choice of representation. So we see that the solution $d$ (and, therefore, $t$ in \eqnrefs{eqn:a1}), $||\fb {p}||$ and $\alpha$ of \eqnref{eqn:a4} will be independent of the representation but $\fb {p}$ in  \eqnrefs{eqn:a1} will depend on the representation.

From \eqnref{eqn:a4} we get : 
\shrteqn{
\alpha = d G_s^{-1} a + G_s^{-1} b.
\label{eqn:a5}
}
From \eqnref{eqn:a2} and \eqnref{eqn:a5}, we obtain :
\shrteqn{
\begin{array}{lll}
p^2 &=& \alpha^T G_s \alpha \\
&=& [a^T(G_s)^{-1}a] d^2 + 2 [b^T (G_s)^{-1}a] d + b^T (G_s)^{-1} b
\end{array}
}
since $G_s$ is symmetric. Equating this to $d^2$ finally gives us a quadratic equation for $d$ : 
\shrteqn{
[a^T(G_s)^{-1}a - 1] d^2 + 2 [b^T (G_s)^{-1}a] d + b^T (G_s)^{-1} b = 0.
\label{eqn:a6}
}
The coefficients of the above quadratic equation are determined solely by the delays $d_i$, $d_{ij}$, and do not depend on the representation. So, if \eqnref{eqn:a6} has a solution $d>0$, our \un{supposition} that the signal left P at some time $t$ and arrived at $s_0$ at the later observed time $t_0$ is a \un{possible} one. Note that \eqnref{eqn:a6} is only a consequence of  \eqnrefs{eqn:a1} and not equivalent to it. Therefore, it does not follow that a solution of \eqnref{eqn:a6} will be a solution of \eqnrefs{eqn:a1}. Since the matrix $G_s$ is positive definite, the constant term in the quadratic is positive if $b \ne 0$. (If $b=0$, then \eqnref{eqn:a6} becomes trivial.)

Now, \un{in principle}, the following cases arise.

\un{Case I} : If $[a^T(G_s)^{-1}a - 1]<0$, then of the two solutions of \eqnref{eqn:a6}, one is positive and the other negative, and so we obtain a unique positive solution of \eqnref{eqn:a6}. Of course, this positive solution may \un{not} give a $t$, and $p$ such that \eqnrefs{eqn:a1} is satisfied. (Using counterexample 1 in Sec. 3.4, with A, B, C, D chosen as $s_0$, $s_1$, $s_2$, $s_3$, E as P, with $a=1.5$, $d_1=d_2=d_3=1$, $d_{12}=d_{23}=d_{31}=\sqrt{2}$, $t_0-t=1.5$, $t_i-d_i-t_0=\sqrt{1+a^2}+1$, so that $t_i-d_i-t_0=0.3028$,  so $[a^T(G_s)^{-1}a - 1]=-0.7249<0$. The positive solution of \eqnref{eqn:a6}, $d=0.5164$ gives $\alpha_1=\alpha_2=\alpha_3=0.2979$ but that does not give $EA=1.5$.) Could a situation arise \un{in practice} where an event cannot be located in $\mathbb{R}^3$, even approximately? In theory, we cannot rule out such a possibility. 

\un{Case II} : If  $[a^T(G_s)^{-1}a - 1]=0$, then there is a unique solution but it may be negative (counterexample 1 with $a=\frac{1}{\sqrt{3}}$.) 

\un{Case III} : If $[a^T(G_s)^{-1}a - 1]>0$, then three cases arise. 

\un{Case III(a)} :  \eqnref{eqn:a6} may have complex solutions in which case \eqnrefs{eqn:a1} has no solutions (counterexample 1 with the following changes: A:$(0,0,0,0)$, B:$(\sqrt{2.61},0,0,0)$, C:$(0,\sqrt{2.61},0,0)$, D:$(0,0,\sqrt{0.99},0)$, E:$(0,0,\sqrt{0.99},0.1)$, $\rm{EA}=1$, $\rm{EB}=\rm{EC}=1.9$, $\rm{ED}=0.1$.) 

\un{Case III(b)} : If the solutions of \eqnref{eqn:a6} are real and both negative (counterexample 1 with $a\le\frac{1}{\sqrt{3}}$), then again \eqnrefs{eqn:a1} has no solution. 

\un{Case III(c)} : Finally, in the third case, \eqnref{eqn:a6} may have two unequal positive solutions.
(Counterexample 2 in Sec. 3.4 gives rise to this case.)
 It is because of this ambiguity, which cannot be resolved in general by using finitely many additional reflecting stations, that we may have to use an echo from the remote object. Thus, the observer would send a signal at time $t'_0$, say, to find out that its echo arrives at the same instant $t_0$ at which the signal possibly generated by the event also arrives and so the time $t$ of the event is immediately determined, ``by definition'', as Einstein says, to be 
\[
t = \frac{1}{2}\ (t'_0+t_0).
\]
In that case, $t$ in \eqnrefs{eqn:a1} is known, and the vector $\fb{p}$ can be uniquely determined from  \eqnref{eqn:a5} itself, provided a solution exists, without going to the quadratic equation. However, it may not be a solution of \eqnrefs{eqn:a1}. It is necessary to check that this solution $\fb{p}$ satisfies the first equation in \eqnrefs{eqn:a1}, namely, $||\fb{p}|| = t_0 - t$. Henceforth, we will assume that the observer does obtain an echo from the remote object, i.e., the observer sends a signal at time $t'_0$ and receives it at time $t_0$.

We have thus shown that the time of occurrence and place (co-ordinates) of a remote event may be \un{defined} and \un{determined} by one observer with one clock, with the help of three reflecting stations. These observations involve only transmission of signals by the observer and reception of signals, direct and indirect from the remote event by the observer. We assume that the observer ascertains through various ``echo'' measurements that the ``distances'' between the stations remain constant. We next turn to the possibility of envisaging another observer and system of observation. 

\section{ANOTHER OBSERVER}

Like Einstein, we now consider the possibility of another observer, or observation system, exactly like the one in the previous section. This second system $\Sigma$ will thus consist of an observer $\sigma_0$, with his own clock, and three reflecting stations $\sigma_1$, $\sigma_2$, $\sigma_3$. $\Sigma$ can also be allowed to assume that the ``velocity'' of the signal is ``1'' but this is a matter of choice. We do choose it to be 1. We will assume that, like the first observer, by ``echo'' experiments, $\sigma_0$ can ascertain that the distances between the various stations do not change with ($\sigma_0$'s) time. Let these distances (and delays) as observed by $\Sigma$ be denoted by $\delta_1$, etc., $\delta_{12}$ etc. We also assume that $\Sigma$ finds $\sigma_0\sigma_1\sigma_2\sigma_3$ to be a non-degenerate tetrahedron so that $\sigma_0$ can determine the time of occurrence $\tau$ and place (co-ordinates) $(\xi,\eta,\zeta)$ of a remote event P on the basis of observation times $\tau'_0$ of sending of the signal, and  times $\tau_0$, $\tau_1$, $\tau_2$, $\tau_3$ of reception of the various echoes. We will find it advantageous now to use a vector $\fb {\pi}$ in place of a triple $(\xi,\eta,\zeta)$, and we will assume that $\fb {\pi}$ belongs to a 3-dimensional inner product space $\Lambda$. Note that this vector space $\Lambda$ need not be the same as the vector space L of the first observer, although since both of them are 3-dimensional, they are isomorphic to one another. 
Let $\Sigma$ assign the zero vector in $\Lambda$ to $\sigma_0$ and vectors $\fb{\delta_1}$, $\fb{\delta_2}$, $\fb{\delta_3}$ in $\Lambda$ to $\sigma_1$, $\sigma_2$, $\sigma_3$, so that $||\fb{\delta_1}||$, $||\fb{\delta_2}||$, $||\fb{\delta_3}||$, $||\fb{\delta_1}-\fb{\delta_2}||$, $||\fb{\delta_2}-\fb{\delta_3}||$, $||\fb{\delta_3}-\fb{\delta_1}||$, are the various delays observed by $\Sigma$. 

\subsection{Relation between the ``Clocks'' of the two Observers}
Now, ``we'' cannot expect to be able to talk about or discover any relation between the determination $t$, $\fb {p}$ by the first observer and the determination $\tau$, $\fb {\pi}$ by the second observer of the ``same'' event P, unless ``we'' assume or discover some relationship between the systems S and $\Sigma$ themselves. So, with Einstein, we assume that $\Sigma$ has a ``uniform motion of translation relatively to'' S. This is something which S (or its observer $s_0$) can ascertain experimentally, and results in a description of the motion of $\Sigma$, i.e., of its stations $\sigma_0$, $\sigma_1$, $\sigma_2$, $\sigma_3$ in S's system. So let the motions of these be given by 
\shrteqn{
\fb {\sigma_0}(t)=\fb {\sigma_0}+t\fb{v},\ \ \ 
\fb {\sigma_i}(t)=\fb {\sigma_0}(t)+\fb {d_i},\ \ \ \ (i=1,2,3)
\label{eqn:a7}
}
where $\fb {\sigma_0}(t)$, $\fb {\sigma_i}(t)$, $\fb {\sigma_0}$, $\fb {d_i}$, and $\fb {v}$ are all vectors, $\fb {d_i}$ being the position vector of $ \sigma_i$ with respect to $\sigma_0$ in S and $\fb {v}$ the common velocity of the stations $\sigma_0$, $\sigma_1$, $\sigma_2$, $\sigma_3$ in S. (The symbols $d_i$, $d_{ij}$ will no longer denote the distances between the stations $s_i$ of S.) 

Thus, we assume that S has ascertained experimentally that the stations $\sigma_0$, $\sigma_1$, $\sigma_2$, $\sigma_3$ of $\Sigma$ form a ``rigid'' system having a common uniform motion of translation relative to S. Now, for $\Sigma$ to be able to assign times and co-ordinates to events in the same manner as S does, it is necessary that these stations form a ``rigid'' system in $\Sigma$, i.e., be at constant $\Sigma$-delays from one another and to form a non-degenerate tetrahedron. Does the rigidity of $\sigma_0\sigma_1\sigma_2\sigma_3$ in S imply the rigidity of $\sigma_0\sigma_1\sigma_2\sigma_3$ in $\Sigma$? Unfortunately, the answer to this is ``no''. We will see below that even \un{assuming} the rigidity of $\sigma_0\sigma_1\sigma_2\sigma_3$ in $\Sigma$ is not enough. However, we do show that assuming the rigidity of the \un{straight line} through $\sigma_0$ and $\sigma_1$ in $\Sigma$ is enough to guarantee the rigidity of $\sigma_0\sigma_1\sigma_2\sigma_3$ in $\Sigma$. In fact, we show that the assumption that the straight line through $\sigma_0$ and $\sigma_1$ is ``rigid'' in $\Sigma$ has the consequence that the \un{$\Sigma$-time} at $\sigma_0$ must be a constant multiple of the \un{S-time} at $\sigma_0$. This relation does hold when $x=vt$ for Einstein's formula 
\[
\tau = \phi(v)\beta(t-vx/c^2)
\]
since with $x=vt$, 
\[
\tau = \phi(v)\beta(1-v^2/c^2)t.
\]
Here, $\phi$ is Einstein's ``yet unknown'' function and $\beta={1}/{\sqrt{1-v^2/c^2}}$. However, in his derivation, Einstein uses the assumption that ``the equations must be \emph{linear} on account of the properties of homogeneity which we attribute to space and time.'' In fact, the linear relation between the $\Sigma$-time $\tau$ and S-time $t$ at $\sigma_0$ follows directly from his assumptions of the linearity of $\tau$ as a function of $x'$, $y$, $z$, $t$, since at $\sigma_0$, $x'=y=z=0$.

Consider \figref{fig:fig1}. (We show a figure only to help ``visualize'' the derivations; we are \un{not} using any ``geometry'' other than that of an inner product space. Interestingly, there were no figures in Einstein's paper, and no references either. However, he does mention Lorentz in one place: \S 9 ``\ldots on the basis of our kinematical principles, the electrodynamic foundation of Lorentz's theory of the electrodynamics of moving bodies is in agreement with the principle of relativity''.)
 \figref{fig:fig1} shows the ``motion'' \un{in S} of a signal which starts from $\sigma_0$ at A at some time $t$, reaches $\sigma_1$ at B at some later time $(t + \Delta'_1)$, and returns to $\sigma_0$ at C at a still later time $(t + \Delta'_1 + \Delta''_1)$.
Here, D is the position of $\sigma_1$ at time $t$, and so $\rm{AD} = d_1$.  E is the position of $\sigma_0$ at time $(t + \Delta'_1)$, so $\rm{EB} = d_1$ also, and EB is parallel to AD.  
We have for the vectors $\fb{\rm{AB}}$, $\fb{\rm{AD}}$, $\fb{\rm{DB}}$,
\[
\begin{array}{lll}
\fb{\rm{AB}}& = & \fb{\rm{AD}} + \fb{\rm{DB}} \\
&=&\fb {d_1}+\Delta'_1 \fb {v}.
\end{array}
\]
Since $||\fb{\rm{AB}}||=\Delta'_1$, we have thus to solve the following equation for $\Delta'_1$:
\shrteqn{
||\fb {d_1} + \Delta'_1\fb {v}|| = \Delta'_1 > 0.
\label{eqn:a8a}
}
We will see that such an equation will occur in our later investigations.

``Squaring'' \eqnref{eqn:a8a},  since $||\fb{\rm{AB}}|| = \Delta'_1$, we get:
\[
{\Delta'_1}^2 = d_1^2 + {\Delta'_1}^2 v^2 + 2(\fb {d_1}\cdot \fb {v}) \Delta'_1
\]
and so:
\shrteqn{
(1-v^2){\Delta'_1}^2 - 2(\fb {d_1}\cdot \fb {v}) \Delta'_1- d_1^2=0.
\label{eqn:new8}
}
\un{Case I} : If $v^2<1$, i.e., the second observer moves at a speed less than that of light, the product of the two roots of this quadratic equation is negative; therefore, it has two real roots, one positive and the other negative. We are assuming a ``direction'' for time, so $\Delta'_1\ge 0$. Thus, it \un{is} possible for the signal leaving $\sigma_0$ at time $t$ to reach $\sigma_1$ at a \un{later} time $(t+\Delta'_1)$, $\Delta'_1$ being the positive root of the quadratic. 

Next, for the vectors $\fb{\rm{BC}}$, $\fb{\rm{BE}}$, $\fb{\rm{EC}}$,
\[
\begin{array}{lll}
\fb{\rm{BC}}& = & \fb{\rm{BE}} + \fb{\rm{EC}} \\
&=&-\fb {d_1}+\Delta''_1 \fb {v}.
\end{array}
\]
Squaring this equation and using $||\fb{\rm{BC}}|| = \Delta''_1$ we obtain a quadratic for $\Delta''_1$:
\shrteqn{
(1-v^2){\Delta''_1}^2 + 2(\fb {d_1}\cdot \fb {v}) \Delta''_1- d_1^2=0.
\label{eqn:new9}
}
Since $v^2<1$, this quadratic, too, has two real roots, one positive and the other negative, so that it \un{is} possible for the signal leaving $\sigma_1$ at time $t+\Delta'_1$ to reach $\sigma_0$ at a \un{later} time $(t+\Delta'_1+\Delta''_1)$, $\Delta''_1$ being the positive root of the quadratic. 

Thus, if $v^2<1$, it is possible for a signal to go from $\sigma_0$ to $\sigma_1$ and then to return to $\sigma_0$, so that $\sigma_0$ will be able to ``see'' $\sigma_1$. (In fact, it is easy to see that the roots of the two quadratic equations are negatives of one another, so that the negative root of the first quadratic could have been interpreted as $-\Delta''_1$.) So, the ``round-trip'' time $\Delta_1$ is given by 
\shrteqn{
\Delta_1 = \Delta'_1 + \Delta''_1 = \frac{2}{1-v^2}\ \sqrt{(1-v^2)d_1^2+(\fb {d_1}\cdot \fb {v})^2},
}
which is independent of the time instant $t$.
 We now assume that $\sigma_0$'s clock shows a $\Sigma$-time $\tau$ which is some function $\phi$ of the S-time $t$ at $\sigma_0$:
\[
 \tau = \phi(t).
\]
This is a \un{special case} of Einstein's assumption that ``to any system of values $x$, $y$, $z$, $t$, which completely defines the place and time of an event in the stationary system (K), there belongs a system of values $\xi$, $\eta$, $\zeta$, $\tau$, determining that event relatively to the system $k$''.

  Now, according to $\Sigma$, $\sigma_1$ is at a fixed ``distance'' from $\sigma_0$, i.e., the round-trip delay from $\sigma_0$ to $\sigma_1$ and back to $\sigma_0$ is constant. So, for all $t$,
\shrteqn{
       \phi(t + \Delta_1) - \phi(t) = k_1,
}
where $k_1$ is some constant, not of our choosing. What kind of a function can we reasonably assume $\phi$ to be ? Surely, if the second observer's ``recording device'' is to deserve the name ``clock'', we expect its time-order to correspond to that of the first observer. (Of course, if the time-order of $\Sigma$ is just the reverse of that of S, we can just change the sign of its reading to restore the correct order.) Also, we would expect $\tau$ to change if $t$ changes. So, the function $\phi$ must be a monotone increasing function. Further, we can ``adjust'' the zero-setting of $\Sigma$'s clock so that $\phi(0)=0$. 

Now, although the linear function 
\[
\theta(t) = (k_1/\Delta_1)t
\]
does satisfy all these conditions, it is not the only function to do so. Indeed, if we let $\psi=\phi-\theta$, we get
\[
\psi(t+\Delta_1) - \psi(t) = 0,\ \ \ \psi(0) = 0,
\]
i.e., that $\psi$ must be a periodic function, with period $\Delta_1$. So, we need some additional conditions on $\phi$ to ``pin it down''. 

We could now invoke the assumption that the other two stations $\sigma_2$, $\sigma_3$, which are at a constant delay from $\sigma_0$ in S, are also at a constant delay from $\sigma_0$ in $\Sigma$, and thus get two additional conditions 
\[
  \phi(t + \Delta_2) - \phi(t) = k_2,\ \ \ 
  \phi(t + \Delta_3) - \phi(t) = k_3,
\]
where $\Delta_2$, $\Delta_3$ are given by expressions similar to the one for $\Delta_1$ and $k_2$, $k_3$ are some further constants, not of our choosing. But, it may happen that, unfortunately, $\Delta_2$, $\Delta_3$ are both equal to $\Delta_1$! Even if they are different from $\Delta_1$, we only get two more conditions on the periodicity of $\psi$, namely that the ``undetermined'' function $\psi$ must be periodic with periods $\Delta_2$, $\Delta_3$ also, provided the constants $k_2$ and $k_3$ are related to $k_1$ ``properly'', namely : 
\[
\frac{k_1}{\Delta_1}= \frac{k_2}{\Delta_2}= \frac{k_3}{\Delta_3},
\]
which could be interpreted as ``isotropy of space''. However, can we conclude from the facts that the function $\psi$ has three different periodicities $\Delta_1$, $\Delta_2$, $\Delta_3$ that $\psi$ must be the zero function? We can if we assume that at least two of these three periods are not rationally related and that the function $\psi$ has a unique Fourier series expansion. But what \un{physical} significance would this assumption have? 

A better alternative, which could be interpreted as homogeneity of space in one direction, is to consider points in S that lie on the infinite straight line through $\sigma_0$ and $\sigma_1$ (somewhat like a co-ordinate axis) and move along with $\sigma_0$ and $\sigma_1$, and to assume that each of these lies at a constant $\Sigma$-delay from $\sigma_0$, proportional to the S-delay from $\sigma_0$. This assumption, together with the fact that $\Delta_1$ is a homogeneous function of $d_1$, would imply that 
\[
\psi(t+\alpha\Delta_1) - \psi(t) = 0,
\]
for every $\alpha$, so $\psi(t)=0$ for all $t$ and thus :
\[
\tau = \beta_1 t,
\]
for some ``yet unknown constant'' $\beta_1$.

Note that our assumption above is not quite what is usually understood as ``rigidity''. Usually, ``rigidity'' is assumed to consist in the invariance of distance between points on a moving object no matter how the object moves. What we have assumed is that points on the $\sigma_0$-$\sigma_1$ ``axis'' which move along with $\sigma_0$ and $\sigma_1$ in S, and, therefore, remain at a constant S-delay from $\sigma_0$, also remain at a constant $\Sigma$-delay. With this assumption, we do not need to assume that $\phi$ is monotone since
\[
\phi(t + \alpha\Delta_1) - \phi(t) = \alpha k_1,
\]
for all $t$ and for all $\alpha$ implies 
\[
\psi(t+\alpha\Delta_1) - \psi(t) = 0,
\]
for all $t$ and for all $\alpha$; this along with $\psi(0)=0$ implies that $\psi(t)=0$ for all $t$.

An advantage of our assumption is that it implies that the stations $\sigma_2$, $\sigma_3$, which move with $\sigma_0$ and $\sigma_1$ in S, stay at constant delays from $\sigma_0$ in $\Sigma$. We will show later (Sec. 5.3) that the stations $\sigma_i$ and $\sigma_j$ also stay at a constant delay from one another in $\Sigma$. So, we need not \un{assume} that $\sigma_0\sigma_1\sigma_2\sigma_3$ is a non-degenerate tetrahedron in $\Sigma$;  this will follow from the fact that $\sigma_0\sigma_1\sigma_2\sigma_3$ is a non-degenerate tetrahedron in S.

  We could now assume that $\beta_1 = 1$, rather than conclude, with Einstein, that  $\beta_1 = \beta =  {1}/{\sqrt{(1-v^2/c^2)}}$, because at this point, we do not see any dependence of $\beta_1$, $\beta_2$ on $v$. We will see now that $\beta_1$ is involved in a relation between the S-distance $d_1$ between $\sigma_0$ and $\sigma_1$, and the $\Sigma$-distance $\delta_1$ between $\sigma_0$ and $\sigma_1$. As seen above, the round-trip S-delay between $\sigma_0$ and $\sigma_1$ is $\Delta_1$ and so the round-trip $\Sigma$-delay between $\sigma_0$ and $\sigma_1$ will be $\beta_1\Delta_1$, and this must equal twice the $\Sigma$-distance $\delta_1$ between $\sigma_0$ and $\sigma_1$; so $2\delta_1=\beta_1\Delta_1$
and thus
\shrteqn{
\begin{array}{lll}
\delta_1=\gamma \sqrt{(1-v^2)d_1^2+(\fb {d_1}\cdot \fb {v})^2}.
\end{array}
\label{eqn:a8}
}
where $\gamma$ denotes ${\beta_1}/{1-v^2}$. 
 \eqnref{eqn:a8} gives a relation between the transition time (in S) $d_1$ from $\sigma_1$ to $\sigma_0$ and the transition time (in $\Sigma$) $\delta_1$ from $\sigma_1$ to $\sigma_0$. We will have similar expressions for $\delta_2$ and $\delta_3$. So, if $\fb{\delta_1}$, $\fb{\delta_2}$, $\fb{\delta_3}$ are vectors in $\Lambda$ which represent $\sigma_1$, 
$\sigma_2$, $\sigma_3$ in $\Sigma$, they must be chosen such that 
\[
||\fb{\delta_i}|| = \delta_i.
\]

\un{Case II} : What would happen if $v^2>1$, i.e., the second observer travels faster than light? The roots of \eqnref{eqn:new8} and \eqnref{eqn:new9} are real only if $|(\fb {d_1} \cdot \fb {v})| \ge 2 d_1 \sqrt{v^2-1}$, and under that condition, since their product is positive, the roots are either both positive or both negative. If, further, $(\fb {d_1} \cdot \fb {v})>0$, the roots of \eqnref{eqn:new8} are both negative. Thus, a signal from $\sigma_0$ cannot reach $\sigma_1$. On the other hand, if $(\fb {d_1} \cdot \fb {v})<0$, then the roots of \eqnref{eqn:new9} are both negative. Thus, a signal from $\sigma_1$ cannot reach $\sigma_0$. Therefore, in either case $\sigma_0$ cannot see $\sigma_1$!  We will comment on the significance of this situation in Sec. 6.3.

\un{Case III} : If $v^2 = 1$, i.e., the second observer travels at the speed of light, only one of   \eqnref{eqn:new8} and \eqnref{eqn:new9} has a positive solution. So, either $\sigma_0$ will not see $\sigma_1$ or $\sigma_1$ will not see $\sigma_0$.

\subsection{Relation between the ``Times'' of a remote Event Determined by the two Observers}
Now we proceed to show that indeed $\fb{\pi}$ and $\tau$ are related to $\fb{p}$ and $t$, and that the relations \un{are} linear. We do not have to \un{assume} like Einstein that ``the equations must be linear on account of the properties of homogeneity which we attribute to space and time''. We have already derived above the linearity of the relation between the S-times and $\Sigma$-times \un{at} $\sigma_0$. We show that $\tau$ and $t$ are also linearly related. 

As remarked above, we will use a vectorial representation and we emphasize that that the two vector spaces used by S and $\Sigma$, namely, L and $\Lambda$, need not be identical. We also find it advantageous to work with the ``relative'' position of P, relative to $\sigma_0$ in S, namely 
\shrteqn{
\fb{\bar{p}}=\fb{p}-\fb{\sigma_0}(t),
}
and thus investigate a possible relationship between $\fb{\bar{p}}$, $t$ and $\fb{\pi}$, $\tau$. 

Firstly, suppose that a signal leaves P, with co-ordinate vector $\fb{p}$ (in S), at time $t$ (in S) and arrives, at some time $t_0$ (in S), at $\sigma_0$, with co-ordinate vector $\fb{\sigma_0}(t_0)$ (in S). (See \figref{fig:fig2}.) Then we must have :
\shrteqn{
||\fb{p}-\fb{\sigma_0}(t_0)|| = t_0 - t.
\label{eqn:anew11}
}
But from \eqnref{eqn:a7}, we have: 
\[
\fb{\sigma_0}(t_0)= \fb{\sigma_0}(t)+(t_0 - t)\fb{v},
\]
so that, 
\shrteqn{
||\fb{\bar{p}}- (t_0 - t)\fb{v}|| = t_0 - t.
\label{eqn:a9}
}
This equation is like  \eqnref{eqn:a8a}. So analogous to \eqnref{eqn:new8}, we obtain :
\shrteqn{
(t_0-t)^2 + \frac{2(\fb{\bar{p}}\cdot \fb{v})}{1-v^2} (t_0-t) - \frac{\bar{p}^2}{1-v^2} = 0.
\label{eqn:a13}
}

Note that the solution of \eqnref{eqn:anew11} for the unknown $t_0$ (and of the equation with $(t-t'_0)$ in place of $(t_0-t)$ in \eqnref{eqn:anew11}, for the unknown $t'_0$) is relatively easy and explicit when the second observer $\sigma_0$ has a uniform motion, as here. When the motion of $\sigma_0$ is not uniform, the solution may be a substantial problem. The solution may not exist, as is the case here if $v^2\ge 1$.

Now, with $v^2<1$, this quadratic equation for $(t_0-t)$ will have two real solutions, one positive and the other negative (because the product of the two roots is negative.) The positive root $(t_0-t)$ (and so with $t_0>t$) gives the time of arrival at $\sigma_0$ of the signal from P, whereas the negative root $(t'_0-t)$ (and so with $t'_0<t$) gives the time of departure from $\sigma_0$ of a signal of which the received signal could well have been an echo from P. 
(If $v^2 \ge 1$, then $\sigma_0$ will not ``see'' P. This is similar to the situation with regard to $\sigma_1$ discussed above.)
The \un{sum} of the two roots of the quadratic is 
\[
-2(\fb{\bar{p}}\cdot \fb{v})/(1-v^2),
\]
and so we obtain 
\[
(t_0 - t) + (t'_0 - t)= -2(\fb{\bar{p}}\cdot \fb{v})/(1-v^2),
\]
giving 
\[
1/2(t_0+t'_0) = t -(\fb{\bar{p}}\cdot \fb{v})/(1-v^2).
\]
But then $\beta_1 t_0$ and $\beta_1 t'_0$ would be the $\Sigma$-times of arrival and departure from $\sigma_0$ of the signal to P, and so we have immediately the time $\tau$ assigned by $\Sigma$ to P as 
\shrteqn{
\begin{array}{lll}
\tau & = & \beta_1[\frac{1}{2}\ ( t_0 +  t'_0)]\\
& = & \beta_1 t - \gamma \ (\fb{\bar{p}}\cdot \fb{v}).
\end{array}
\label{eqn:a21}
}
Note that essentially what we have derived in this sub-section is a formula for the time $t'_0$ in S of the departure of the signal from the station $\sigma_0$ to the observed event ($p$, $t$) and the time $t_0$ in S of the arrival of this signal at $\sigma_0$. Note also that in this derivation, we have not made any assumption about the speed of light in the moving systems $\Sigma$.
 
\subsection{Relation between the ``Co-ordinates'' of a remote Event Determined by the two Observers}
Now, in $\Sigma$, the same signal (see \figref{fig:fig2}) leaves from an (unknown) place $\fb{\pi}$ at the $\Sigma$-time $\tau$ that we have calculated, and arrives at $\sigma_0$ at S-time $t_0$, and so, at $\Sigma$-time $\beta_1 t_0$, and thus we have
\shrteqn{
\begin{array}{lll}
||\fb{\pi}||&=&\beta_1 t_0 - \tau\\
&=&\beta_1(t_0-t)+\gamma(\fb{\bar{p}}\cdot \fb{v}). 
\end{array}
\label{eqn:a11}
}

From \eqnref{eqn:a13}, we obtain
\[
\begin{array}{lll}
t_0-t&=&\frac{1}{2}\ \left [\  -2 \frac{\fb{\bar{p}}\cdot \fb{v}}{1-v^2}+\sqrt{4\frac{(\fb{\bar{p}}\cdot \fb{v})^2}{(1-v^2)^2}+4\frac{\bar{p}^2}{(1-v^2)}}\ \right ]\\
&&\\
&=&\frac{1}{1-v^2}\ \left [\ \sqrt{(\fb{\bar{p}}\cdot \fb{v})^2+(1-v^2)\bar{p}^2}-\fb{\bar{p}}\cdot \fb{v}\ \right ]
\end{array}
\]
and so we get
\shrteqn{
||\fb{\pi}|| = \gamma\sqrt{(\fb{\bar{p}}\cdot \fb{v})^2+(1-v^2)\bar{p}^2}.
}
Squaring both sides, we get
\shrteqn{
\pi^2 = \gamma^2 \ \left [\ (\fb{\bar{p}} \cdot \fb{v})^2 + (1-v^2)\bar{p}^2\ \right ].
\label{eqn:a22}
}
Similarly, suppose the signal, leaving P at $t$, arrives at $\sigma_0$ at S-time $t_1$ via $\sigma_1$. (See \figref{fig:fig2}.) It must have been at $\sigma_1$ at S-time $t_1-\Delta''_1$. So 
\[
||\fb{p}-\fb{\sigma_1}(t_1-\Delta''_1)|| = t_1-\Delta''_1 - t.
\]
But since
\[
\fb{\sigma_1}(t)=\fb{\sigma_0}(t)+\fb{d_1}
\]
so
\[
\fb{p}-\fb{\sigma_1}(t_1-\Delta''_1)=\fb{\bar{p}}-\fb{d_1}-(t_1-\Delta''_1 - t)\fb{v}
\]
and thus,
\shrteqn{
||\fb{\bar{p}}-\fb{d_1}-(t_1-\Delta''_1 - t)\fb{v}|| = t_1-\Delta''_1 - t.
\label{eqn:a10}
}
From \eqnref{eqn:a10}, we  obtain $(t_1-\Delta''_1 - t)$, just as from \eqnref{eqn:a9}, we obtained $(t_0-t)$:
\[
t_1-\Delta''_1 - t = \frac{1}{1-v^2}\ \left [ \sqrt{\{(\fb{\bar{p}} - \fb{d_1})\cdot \fb{v}\}^2 + (1-v^2)(\fb{\bar{p}}-\fb{d_1})^2}-(\fb{\bar{p}}-\fb{d_1})\cdot \fb{v}\ \right ].
\]
 
This signal arrives at $\sigma_0$ via $\sigma_1$ at S-time $t_1$, and so, at $\Sigma$-time $\beta_1 t_1$. (See \figref{fig:fig3}.) So, $\Sigma$ will assign to the signal the time ($\beta_1 t_1 - \delta_1$) of arrival at $\sigma_1$, and because
\[
\delta_1=\beta_1[\Delta_1''+(\fb{d_1}\cdot \fb{v})/(1-v^2)],
\]
so, 
\shrteqn{
\begin{array}{lll}
||\fb{\pi}-\fb{\delta_1}|| &= &(\beta_1 t_1 - \delta_1) - \tau\\
&=&\beta_1 (t_1 - \Delta''_1-t)+\gamma\{(\fb{\bar{p}}- \fb{d_1})\cdot \fb{v}\}\\
&=&\gamma \left [\  \sqrt{\{(\fb{\bar{p}} - \fb{d_1})\cdot \fb{v}\}^2 + (1-v^2)(\fb{\bar{p}}-\fb{d_1})^2}\ \right ].
\end{array}
\label{eqn:a12}
}
(We could have obtained \eqnref{eqn:a12} directly by putting $\fb{\pi} - \fb{\delta_1}$ in place of $\fb{\pi}$ and $\fb{\bar{p}}-\fb{d_1}$  in place of $\fb{d_1}$ in \eqnref{eqn:a22}.)

Squaring both sides of \eqnref{eqn:a12}, since $||\fb{\delta_1}|| = \delta_1$, we get
\[
\pi^2-2\fb{\pi} \cdot \fb{\delta_1} + \delta_1^2 = \gamma^2 \ \left [\ \{(\fb{\bar{p}} - \fb{d_1})\cdot \fb{v}\}^2 + (1-v^2)(\fb{\bar{p}}-\fb{d_1})^2 \ \right ].
\]
Using \eqnref{eqn:a22} and \eqnref{eqn:a8}, this gives
\shrteqn{
\fb{\pi}\cdot \fb{\delta_1} = \gamma^2\ \left [\ (1-v^2)( \fb{\bar{p}}\cdot \fb{d_1})-(\fb{\bar{p}}\cdot \fb{v})(\fb{d_1}\cdot \fb{v})\ \right ].
\label{eqn:a17}
}

Now, the right-hand side of \eqnref{eqn:a17} can be written as $\fb{w_1}\cdot\fb{\bar{p}}$ where
\shrteqn{
\fb{w_1}=\gamma^2\ \left [\ (1-v^2)(\fb{d_1}) - (\fb{d_1}\cdot \fb{v})\fb{v}\ \right ]  
}
so that \eqnref{eqn:a17} becomes:
\shrteqn{
\fb{\pi}\cdot\fb{\delta_1}=\fb{w_1}\cdot\fb{\bar{p}}\ .
\label{eqn:a18}
}
We can similarly obtain two more equations:
\shrteqn{
\fb{\pi}\cdot\fb{\delta_2}=\fb{w_2}\cdot\fb{\bar{p}}\ ,
\label{eqn:a19}
}
\shrteqn{
\fb{\pi}\cdot\fb{\delta_3}=\fb{w_3}\cdot\fb{\bar{p}}\ .
\label{eqn:a20}
}

Since $\sigma_0\sigma_1\sigma_2\sigma_3$ is a non-degenerate tetrahedron in $\Sigma$, the vectors $\fb{\delta_1}$, $\fb{\delta_2}$, $\fb{\delta_3}$ are linearly independent, so form a basis for $\Lambda$, and so $\fb{\pi}$ is a linear combination of $\fb{\delta_1}$, $\fb{\delta_2}$, $\fb{\delta_3}$. This, alongwith  \eqnref{eqn:a18}, \eqnref{eqn:a19}, \eqnref{eqn:a20} implies that there must be a linear transformation T from the vector space L to the vector space $\Lambda$:
\[
T:L\rightarrow\Lambda
\]
such that
\shrteqn{
\fb{\pi}=T(\fb{\bar{p}}). 
}

Thus, we have \un{proved} that ${\pi}$ and $\tau$ are linearly related to $\fb{\bar{p}}$, \un{the relation being independent of $t$}. The transformation $T$ can be calculated from  \eqnref{eqn:a18}, \eqnref{eqn:a19}, \eqnref{eqn:a20}, but knowing that such a transformation exists, we can take a short-cut. Indeed, choose P to be $\sigma_1$ itself in S. Then, $\fb{\bar{p}} = \fb{d_1}$. But $\sigma_1$ is $\fb{\delta_1}$ in $\Sigma$. So we have
\shrteqn{
T(\fb{d_1})=\fb{\delta_1}.
}
Similarly,
\shrteqn{
T(\fb{d_2})=\fb{\delta_2}, 
}
\shrteqn{
T(\fb{d_3})=\fb{\delta_3}.
}
From this, it follows that the transformation $T$ is one-to-one.

Now, this co-ordinatization in $\Sigma$ involves choosing the representing vectors $\fb{\delta_1}$, $\fb{\delta_2}$, $\fb{\delta_3}$ in $\Lambda$. They cannot be chosen arbitrarily, however, because we have the equations
\[
||\fb{\delta_i}||=\gamma \{ \sqrt{(1-v^2)d_i^2+(\fb{d_i}\cdot\fb {v})^2} \}
\]
obtained earlier. Moreover, by considering signals traveling between, say, $\sigma_1$ and $\sigma_2$ in S and in $\Sigma$, we get additional relations for the norms of the vectors $\fb{\delta_{12}}$, $\fb{\delta_{23}}$ and $\fb{\delta_{31}}$, these vectors forming the remaining three sides of the tetrahedron $\sigma_0\sigma_1\sigma_2\sigma_3$ in $\Sigma$. We can calculate the $\Sigma$-distance $\delta_{12}$ between $\sigma_1$ and $\sigma_2$ as follows (see \figref{fig:fig4}). 

Consider a round-trip from $\sigma_0$ to $\sigma_1$ to $\sigma_2$ back to $\sigma_0$, starting at $\sigma_0$ at A at S-time $t$. This signal will reach $\sigma_1$ at B at S-time 
\[
t+\Delta'_{01}
\]
where, 
\[
\Delta'_{01}=\frac{1}{1-v^2} \{ (\fb{d_1} \cdot \fb{v}) + \sqrt{(1-v^2) d_1^2 + (\fb{d_1}\cdot \fb{v})^2} \}.
\]
It will reach $\sigma_2$ at C at a time $\Delta'_{12}$ later where
\[
\Delta'_{12}=\frac{1}{1-v^2} \{ (\fb{d_{12}} \cdot \fb{v}) + \sqrt{(1-v^2) d_{12}^2 + (\fb{d_{12}} \cdot \fb{v})^2} \}.
\]

Finally, it will return to $\sigma_0$ at D at a time $\Delta''_{20}$ later where
\[
\Delta''_{20}=\frac{1}{1-v^2} \{ -(\fb{d_2} \cdot \fb{v}) + \sqrt{(1-v^2) d_2^2 + (\fb{d_2}\cdot \fb{v})^2} \}.
\]
Thus, the round-trip time in S is 
\[
\Delta'_{01}+\Delta'_{12}+\Delta''_{20}
\]
and so, in $\Sigma$, it is 
\[
\beta_1(\Delta'_{01}+\Delta'_{12}+\Delta''_{20}).
\]
From this, we subtract the $\Sigma$-delay $\delta_1$ between $\sigma_0$ and $\sigma_1$ and the delay $\delta_2$ between $\sigma_2$ and $\sigma_0$, to obtain the $\Sigma$-delay $\delta_{12}$ between $\sigma_1$ and $\sigma_2$: 
\[
\begin{array}{lll}
\delta_{12} &=& \gamma \sqrt{(1-v^2) d_{12}^2 + (\fb{d_{12}} \cdot {v})^2}
\end{array}
\]
since 
\[
\fb{d_{12}} + \fb{d_1} - \fb{d_2} = \fb{0}.
\]
Note also that $\delta_{12} = \delta_{21}$.

We now show that the six delays, $\delta_1$, $\delta_2$, $\delta_3$, $\delta_{12}$, $\delta_{23}$, $\delta_{31}$, satisfy the appropriate triangle inequalities. Let $\fb{\bar u}$ be a unit vector orthogonal to $\fb{d_1}$ and $\fb{d_2}$. Defining new vectors $\fb{\overline{d_1}}$, $\fb{\overline{d_2}}$, $\fb{\overline{d_{12}}}$ as follows : 
\[
\begin{array}{lll}
\fb{\overline{d_1}} & = & \sqrt{(1-v^2) }\fb{d_1} +       (\fb{d_1} \cdot \fb{v}) \fb{\bar u}, \\
\fb{\overline{d_2}} & = & \sqrt{(1-v^2) }\fb{d_2} +       (\fb{d_2} \cdot \fb{v}) \fb{\bar u}, \\
\fb{\overline{d_{12}}} & = & \sqrt{(1-v^2) }\fb{d_{12}} + (\fb{d_{12}} \cdot \fb{v}) \fb{\bar u},
\end{array}
\]
since $\fb{d_{12}}+\fb{d_1}-\fb{d_2}=0$, we have $\fb{\overline{d_{12}}} + \fb{\overline{d_1}} - \fb{\overline{d_1}} = 0$. Thus, these three vectors form a triangle and therefore, their norms, which are equal to the delays $\delta_1$, $\delta_2$ and $\delta_{12}$, satisfy triangle inequalities.

What we have shown above is that the six sides of the tetrahedron $\sigma_0\sigma_1\sigma_2\sigma_3$ in $\Sigma$ can be calculated from the six sides of the tetrahedron in S and the  vectors $\fb{\sigma_1}$, $\fb{\sigma_2}$, $\fb{\sigma_3}$, and the velocity vector $\fb{v}$ in S. As remarked in Sec. \ref{subsec:3d}, the tetrahedron in $\Sigma$ can be co-ordinatized or represented by vectors in the inner-product space $\Lambda$ non-uniquely. Note that the tetrahedron $\sigma_0\sigma_1\sigma_2\sigma_3$ moves in S but it remains rigid and has a translatory motion. The transformation $T$ that maps the position of the remote event relative to $\sigma_0$ in S into the position of that event relative to $\sigma_0$ in $\Sigma$ is the one that maps the relative position vector $\fb{\sigma_1}-\fb{\sigma_0}$, namely, $\fb{d_1}$, in L into the vector $\fb{\delta_1}$ in $\Lambda$, and so on. 

What about description of motion in S and $\Sigma$? To be able to define velocity and acceleration, the geometry of an inner-product space suffices. But we do not develop this here.

\subsection{Theory and Practice of Observers}
\subsubsection{Could the reflecting stations ``move''?}
It is perhaps too ``theoretical'' or ``idealistic'' to assume that the distances between the various stations remain constant. To be ``practical'', one should consider the possibility of these distances (delays, really) varying with $s_0$'s time (or $\sigma_0$'s time). In such a case, the lengths of the sides of the tetrahedron $s_0s_1s_2s_3$ will change with time. But what do we mean here by ``the sides of the tetrahedron at any instant of time''? We had assumed that the observer, with his clock at $s_0$, finds out by various echo measurements that the delays between the stations remain constant. This allowed us to \un{define} the various distances between the stations. Now, if the delays do not remain constant, then we can still give meaning to the distance between, say, $s_0$ and $s_1$, at each instant of time on the basis of echo measurement, a la Einstein. Thus, if a signal leaves $s_0$ at $t_1$ and returns from $s_1$ at $t_1'$, the distance between $s_0$ and $s_1$ at time ${1}/{2}(t_1+t_1')$ is ${1}/{2}(t_1'-t_1)$. Similarly, for the distance between $s_0$ and $s_2$, and the distance between $s_0$ and $s_3$. But, what does ``distance between $s_1$ and $s_2$'' mean and at what time? The point to be noted is that happenings at the stations $s_1$, $s_2$, $s_3$ cannot be treated as just some events. The stations and observations made on them provide a foundation for \un{defining} and calculating the time and position of a remote event. For this, as pointed out earlier, we need \un{all} the four stations. The position of a reflecting station, say $s_1$, cannot be determined using only the three stations $s_0$, $s_2$ and $s_3$. Perhaps, the only way out is to adjust the positions of $s_1$, $s_2$, $s_3$ so that the tetrahedron $s_0s_1s_2s_3$ remains unchanged. 
If, however, this is not done, the observer could keep sending the signal and keep receiving echoes, three direct and six indirect, from $s_1$, $s_2$, $s_3$, in all possible ways, thereby obtaining nine functions of the time of transmission. To this, he could add four echoes, one direct and three indirect, from the object being observed. Thus, he would have thirteen functions of time as his experimental data for the object being observed. Perhaps, this data could be used to build a model of the motions of the stations and the object being observed.

Similar considerations will apply to the other observer. It may seem that we could do away with the assumption that the stations $\sigma_0$, $\sigma_1$, $\sigma_2$, $\sigma_3$ move together with the same constant velocity if we suppose that system S ascertains their positions continuously. But what about $\sigma_0$'s own observations of $\sigma_1$, $\sigma_2$, $\sigma_3$? We do not want to assume any a priori relation between S's clock and $\Sigma$'s clock. We assumed that there was \un{some} relation between these clocks, and that what S finds to be ``rigid'' (the tetrahedron $\sigma_0\sigma_1\sigma_2\sigma_3$) $\Sigma$ also finds to be rigid.
Perhaps, one could assume that $\Sigma$ borrows S's time, that is to say, that the times at $\sigma_0$ are what S assigns to them. It seems that this is exactly what the GPS receiver does; it updates or corrects its clock on the basis of signals received from the space vehicles.

In practice, in GPS, one does expect that the master control station and the monitor stations do remain fixed relative to one another. (Does one actually check this out?) Of course, in the event of an earthquake (or continental drift), these distances could change with time.

\subsubsection{Is the second observer necessary ?}
Is it necessary to think of the \un{second} observer as a ``real'' observer, with his own ``real'' clock? Could we not let the \un{first} observer perform a ``gedanken'' calculation to find out what a second observer \un{would} observe? (This is precisely what we have done!) As indicated towards the end of Sec. 5.2, for an event P ($\fb{p}$, $t$) observed by the first observer S, assuming that the system $\Sigma$ has a ``known'' uniform motion relative to S, the departure and arrival times $t'_0$, $t_0$, $t_1$, $t_2$, $t_3$ at $\sigma_0$ can be calculated in terms of $\fb{p}$, $t$, and the $\delta$'s. As shown in Sec. 5.1, ``rigidity'' in $\Sigma$ of the straight line through $\sigma_0$ and $\sigma_1$ implies a simple relation between the two clocks, namely : 
\[
\tau = \beta_1 t.
\]
Why not then dispense with $\Sigma$'s clock and let him instead use S's clock, i.e., to use these \un{calculated} time instants? This will do away with the problem of what happens to a ``real'' clock when it is put in uniform motion. Of course, $\Sigma$'s calculation of the space aspect $\fb{\pi}$ of the event P are different from S's calculation $\fb{p}$ simply because $\Sigma$'s data is different. Also, \Si's calculation of the time aspect $\tau$ of the event P is different from S's calculation $t$ for the same reason. 

\section{SOME FURTHER CONSIDERATIONS}
\subsection{Comparison with Einstein's Formulas: Rectangular Cartesian co-ordinate Systems}
In his derivation,Einstein makes \un{use} of the \un{classical} ``relative position'' of an event, relative to a moving observer, namely: 
\[
x' = x - vt
\]
where $x$ is the position of the event in the ``stationary system'' and $v$ is the velocity of the ``moving system''. (His ``$x'$'' thus corresponds to our $\fb{\bar{p}}=\fb{p}-\fb{\sigma_0}(t)$.) However, in the final formulas he uses the variable $x$. His formulas relating the time and place determinations by the two observers are:
\shrteqn{
\begin{array}{lll}
\tau & = &\phi(v)\beta(t-vx/c^2),\\
\xi&=&\phi(v)\beta(x-vt),\\
\eta&=&\phi(v)y,\\
\zeta&=&\phi(v)z,
\end{array}
}
where $\beta={1}/{\sqrt{1-v^2/c^2}}$. Note that $\xi$ differs from $(x-vt)$ by a multiplying coefficient, whereas the expression for $\tau$ involves a peculiar combination of $t$ and $x$. (Einstein's derivation of the formula for $\xi$ appears to be incomplete because he derives it only for a special class of events, namely $x=ct$, $y=z=0$. Similarly, regarding his formulas for $\eta$, $\zeta$.)

Using $x'$, rather than $x$, the formulas become:
\shrteqn{
\begin{array}{lll}
\tau&=&\phi(v)\beta[(1-v^2/c^2)t - (v/c^2)x'],\\
\xi&=&\phi(v)\beta x',\\
\eta&=&\phi(v)y,\\
\zeta&=&\phi(v)z.
\end{array}
}
Using our derivation, we choose (i) the vector spaces L and $\Lambda$ both to be $\mathbb{R}^3$; (ii) $\fb{s_1}$, $\fb{s_2}$, $\fb{s_3}$, the position vectors of the stations $\sigma_1$, $\sigma_2$, $\sigma_3$, with respect to $\sigma_0$ in S to be the three unit vectors $\fb{i}$, $\fb{j}$, $\fb{k}$ in $\mathbb{R}^3$, so that S uses a rectangular Cartesian co-ordinate system, and (iii) choose the velocity $\fb{v}$ to be in the direction of the positive x-axis so that $\fb{v}=||\fb{v}||\fb{i}$. After calculating $||\fb{\delta_1}||$, $||\fb{\delta_2}||$, $||\fb{\delta_3}||$, $||\fb{\delta_{12}}||$, $||\fb{\delta_{23}}||$, $||\fb{\delta_{31}}||$, we see that we can \un{choose} the vectors $\fb{\delta_1}$, $\fb{\delta_2}$, $\fb{\delta_3}$ to be 
\[
\fb{\delta_1}=\frac{\beta_1}{1-v^2}\fb{i},\ \ \fb{\delta_2}=\frac{\beta_1}{\sqrt{1-v^2}}\fb{j},\ \ \fb{\delta_3}=\frac{\beta_1}{\sqrt{1-v^2}}\fb{k},
\]
so that $\Sigma$  also uses a rectangular Cartesian co-ordinate system.

The transformation $T:L\rightarrow\Lambda$ is given by the diagonal matrix 
\[
T=\mbox{diag}\ \left [\ \frac{\beta_1}{1-v^2},\ \frac{\beta_1}{\sqrt{1-v^2}},\ \frac{\beta_1}{\sqrt{1-v^2}}\ \right ]
\]
so that $\fb{\pi} = T\fb{p}$ gives 
\[
\xi = \frac{\beta_1}{1-v^2}x',\ \ 
\eta = \frac{\beta_1}{\sqrt{1-v^2}}y,\ \ 
\zeta = \frac{\beta_1}{\sqrt{1-v^2}}z.
\]
These will agree with Einstein's formulas if we choose $c=1$ and 
\[
\phi(v)={\beta_1}/\sqrt{1-v^2}.
\]
Our formula for $\tau$ then gives 
  
\[
\tau=\beta_1(t-vx'/(1-v^2))
\]
which agrees with Einstein's expression.

Hence, we can say that we have generalized Einstein's derivation in two respects: firstly, we have \un{defined} the co-ordinate system used operationally, and secondly, we have derived the formulas for any arbitrary direction of the velocity $v$ of the ``moving'' system relative to the ``stationary'' system. 

\subsection{Velocity of Light in the two Systems}
Where have we used the velocity of light in our derivations? We have used it in defining ``distance'' in terms of ``time difference'' and we chose it to be ``1'' and the same for both systems. It should be clear that the choice of the velocity of light in each system is arbitrary. We do not imply by this that the velocity of light is arbitrary in practice, or that it cannot be measured in practice. What we mean is that in defining \un{distance} in our approach, a constant is used which, by analogy with mechanics of stretched cords, rigid rods and mobile observers, may be called ``velocity of light'' in our system. It seems to us that the times and distances, and, therefore, also the velocity of light, in the two systems are non-commensurate in the sense of Kuhn. System S cannot \un{measure} velocity of light in its own system, leave alone measure the velocity of light in $\Sigma$'s system! We have, of course, assumed a ``correspondence'' between the clocks of S and $\Sigma$, but this is something which neither S nor $\Sigma$ alone could observe. $\Sigma$'s clock is not observable by S and vice versa. Perhaps, only an omniscient observer can act as a go-between and actually observe that the two times \un{at $\sigma_0$} are related, by conducting an appropriate experiment. Thus, one (or someone on one's behalf - unless one travels \un{with} the light signal) can send a signal from $\sigma_0$ to $s_0$ at observed $\Sigma$-time $\tau_0$, and receive it at $s_0$ at some observed S-time $t$ from which one can calculate the S-time $t_0$ of departure of the signal from $\sigma_0$, and then find that 
\[
\tau_0 = \beta_1 t_0,
\]
where $\beta_1$ is a constant, independent of $\tau_0$.

\subsection{``Faster-than-Light'' Observer, ``Faster-than-light'' Particles and Composition of Velocities }

``Faster - than - Light''  particles have been much discussed in the literature \cite{FAYNGOLD}.  In our approach, we determine the time and place of an event by using the transmission and reception times of signals.  An event is detected or recognized to have happened only if the observer (in S) receives an echo and reflected signals \un{and} these signals satisfy the necessary triangle inequalities for a tetrahedron.  We have, therefore, no basis for answering the question: ``can there be a faster-than-light particle?''.

However, if the observer in S does find that the other system $\Sigma$  is moving faster than light ($||\fb{v}|| \ge 1$), then as pointed out above, $\sigma_0$ of $\Sigma$ will not be able to see any event P that is seen by S, because no echo from P will reach $\sigma_0$. 

We thus conclude that it is useless to consider a system $\Sigma$ which moves, relative to S, faster that light because such a system will not see any event seen by S. Note that the reason for this happening is not any expression  like ``$\sqrt{1-v^2}$'' appearing in our derivation. Einstein had concluded that ``for velocities greater than that of light our deliberations become meaningless'' for a different reason, namely, the shortening of lengths by the factor $\sqrt{1-v^2}$. 

But the above considerations need \un{not} deter us from envisaging faster-than-light \un{particles}, because we have shown above that any event P which is seen by S is also visible to $\Sigma$, provided $||\fb{v}||<1$. Indeed, we can derive a law of ``composition of velocities''. Let P move uniformly with a velocity $\fb{w}$, i.e., we consider a family of events given by 
\[
\fb{p}(t) = t\fb{w} + \fb{\rm{constant}},
\]
and so 
\[
\fb{\bar{p}}(t) = t(\fb{w}-\fb{v}) + \fb{\rm{constant}},
\]
Then, for $\tau$ and $\fb{\pi}$, we obtain:
\[
\begin{array}{lll}
\tau &=& \beta_1 t - \gamma [(\fb{w}-\fb{v})\cdot \fb{v}] \\
&=& \gamma [ 1-(\fb{w}\cdot \fb{v})]t
\end{array}
\]
and 
\[
\fb{\pi} = T(\fb{\bar{p}}) = t T(\fb{w}-\fb{v}) + \fb{\rm{constant}},
\]
so that the velocity of P relative to $\Sigma$ will be:
\[
{T(\fb{w}-\fb{v})}/(\gamma[1-(\fb{w}\cdot \fb{v})]).
\]
This leads to the following possibilities. If $\fb{w}\cdot \fb{v}<1$, which can happen even if $||\fb{w}||>1$, i.e., P travels faster than light, then the ``direction'' of ``time'' $\tau$ in $\Sigma$ is same as the ``direction'' of ``time'' $t$ in S. But if $\fb{w} \cdot \fb{v} > 1$, then there is a time-reversal from S to $\Sigma$. However, this time reversal will not be seen by both S and $\Sigma$; it is only an omniscient observer who will notice it. If $\fb{w}\cdot \fb{v}=1$, then $\tau=0$, so that $\Sigma$ will see P's whole ``history'' in one moment!

In the special case when $\fb{w}$ is in the same  direction as $\fb{v}$, we obtain a simple expression for the magnitude of the relative velocity, which is similar to Einstein's formula. Indeed, if $\fb{w}=k\fb{v}$ where $k$ is a real number, then the magnitude of the relative velocity is:
\[
|\ (k-1)/(1-k v^2)\ |\ ||\fb{v}||.
\]

It is interesting to note that Einstein in his paper has \un{nowhere} ruled out faster-than-light particles. We show in Sec. 7.4 that there is no difficulty in considering faster-than-light particles in electrodynamics.

\subsection{``One-way'' and ``Two-way'' Velocity of Light}
It should be clear that in our approach, velocity of the signal (light) is an ``undefined'' concept. It is more like a mere number, used to define ``distance'' in terms of time. So the velocity of light, whether one-way or two-way, is not something which can be measured in our system. Also, ``homogeneity'' or ``isotropy'' perhaps are not properties of some independently conceived or experienced ``space'', but are rather a matter of assumption about the representation of ``travel'' of light.

\subsection{Symmetry}
In the calculations above, $\fb{v}$ was the velocity of $\Sigma$ as seen by S. What will be the velocity of S as seen by $\Sigma$? Will it be $\fb{-v}$? No, we should not expect it to be $\fb{-v}$ simply because of the choice involved, of the vector spaces L and $\Lambda$, in setting up the co-ordinate systems in S and $\Sigma$. But we can use our formula relating $\fb{\pi}$ and $\fb{\bar{p}}$, by taking P to be the origin of S, so that 
\[
\fb{\bar{p}}(t) = -\fb{\sigma_0}(t) = -\fb{\bar{\sigma_0}}-t\fb{v}
\]
and since 
\[
\begin{array}{lll}
\tau &=& \beta_1 t-\gamma\ (\fb{\bar{p}}\cdot \fb{v})\\
&=&\gamma\ (t + \fb{\bar{\sigma_0}} \cdot {v})
\end{array}
\]
we have in $\Sigma$
\[
\fb{\pi} = T(\fb{\bar{p}}) = -T(\fb{\bar{\sigma_0}})+(\fb{\bar{\sigma_0}} \cdot \fb{v} )T(\fb{v})-(\tau/\gamma)T(\fb{v}) 
\]
so that the velocity of S relative to $\Sigma$ will be 
\[
-(1/\gamma)T(\fb{v})
\]
which need not be $\fb{-v}$. ($\fb{v}$ is in the space L, $T(\fb{v})$ is in the space $\Lambda$.) However, we show below that $||T(\fb{v})||$ is equal to $\gamma ||\fb{v}||$, which means that the \un{magnitude} of the velocity of S in $\Sigma$ is the same as the the magnitude of the velocity of $\Sigma$ in S. In Einstein's special case discussed above, where $\fb{v}=||\fb{v}||\fb{i}$ and L and $\Lambda$ are both $\mathbb{R}^3$, $T(\fb{v})=-\fb{v}$, independent of how $\beta_1$ is chosen! Indeed, in Einstein's formula, the velocity of K (our S) relative to k (our $\Sigma$) is $-v$ independent of how $\phi(v)$ is chosen.(Einstein chooses $\phi(v)=1$, using some symmetry conditions for the motion.)

To show that $||T(\fb{v})|| = \gamma ||\fb{v}||$, we first evaluate the Gram matrix $G_{\delta}$ of the three vectors $\{\fb{\delta_1}, \fb{\delta_2}, \fb{\delta_3}\}$ given by
\[
(G_{\delta})_{ij} = \fb{\delta_i} \cdot \fb{\delta_j}.
\]
Using the formulas for $\delta_i^2$ and $\delta_{ij}^2$ and the definitions
\[
\fb{\delta_{ij}} = \fb{\delta_j}-\fb{\delta_i},\ \ \fb{d_{ij}}=\fb{d_j}-\fb{d_i}
\]
we obtain $\ \fb{\delta_i}\cdot\fb{\delta_j} = (1-v^2)(\fb{d_i}\cdot \fb{d_j})+(\fb{d_i} \cdot \fb{v}) (\fb{d_j} \cdot \fb{v})$. 

So
\[
G_{\delta} = \gamma^2(1-v^2)G_d+\gamma^2
\left [
\begin{array}{lll}
\fb{d_1} \cdot \fb{v}\  & \fb{d_2} \cdot \fb{v}\  & \fb{d_3} \cdot \fb{v} 
\end{array}
\right ]^T
\left [
\begin{array}{lll}
\fb{d_1} \cdot \fb{v} \ & \fb{d_2} \cdot \fb{v}\ & \fb{d_3} \cdot \fb{v} 
\end{array}
\right ]
\]
where $G_d$ is the Gram matrix for the set $\{\fb{d_1}, \fb{d_2}, \fb{d_3}\}$.

Letting $\fb{v}=\alpha_1 \fb{d_1} + \alpha_2 \fb{d_2} +\alpha_3 \fb{d_3}$, we obtain
\[
v^2 = \alpha^T G_d \alpha
\] 
where $\alpha = \left [ \begin{array}{lll} \alpha_1 & \alpha_2 & \alpha_3 \end{array} \right ]^T$. 

We then have
\[
\begin{array}{lll}
T(\fb{v}) &=&  \alpha_1 T(\fb{d_1}) + \alpha_2 T(\fb{d_2}) +\alpha_3 T(\fb{d_3}) \\ 
&=&\alpha_1 \fb{\delta_1} +\alpha_2 \fb{\delta_2} +\alpha_3 \fb{\delta_3} 
\end{array}
\]
so $T(\fb{v})^2 = \alpha^T G_{\delta} \alpha$. 

Now 
\[
\alpha^T G_{\delta} \alpha
\begin{array}{lll}
&=& \gamma^2 (1-v^2) \alpha^T G_d\alpha + \gamma^2\alpha^T
\left [
\begin{array}{lll}
\fb{d_1} \cdot \fb{v} \ & \fb{d_2} \cdot \fb{v} \ & \fb{d_3} \cdot \fb{v} 
\end{array}
\right ]^T
\left [
\begin{array}{lll}
\fb{d_1} \cdot \fb{v} \ & \fb{d_2} \cdot \fb{v} \ & \fb{d_3} \cdot \fb{v} 
\end{array}
\right ]
\alpha
\end{array}
\]
but since
\[
\begin{array}{lll}
\left [
\begin{array}{lll}
\fb{d_1} \cdot \fb{v} \ & \fb{d_2} \cdot \fb{v} \ & \fb{d_3} \cdot \fb{v} 
\end{array}
\right ] \alpha 
&=&\alpha_1(\fb{d_1} \cdot \fb{v}) +\alpha_2(\fb{d_2} \cdot \fb{v}) +\alpha_3(\fb{d_3} \cdot \fb{v}) \\
&=&\fb{v}\cdot (\alpha_1 \fb{d_1} + \alpha_2 \fb{d_2} +\alpha_3 \fb{d_3})\\
&=&v^2,
\end{array}
\]
so 
\[
\begin{array}{lll}
T(\fb{v})^2 &=& \gamma^2 (1-v^2) v^2 + \gamma^2 (v^2)^2\\
&=&\gamma^2 v^2,
\end{array}
\]
and thus $\ \ ||T(\fb{v})|| = \gamma ||\fb{v}||$.

\subsection{The ``Group Property'' and ``Inertial Frames of Reference''}
We first note that Einstein uses the expression ``system of co-ordinates'' rather than ``inertial frame of reference''. We can see from our derivations in Sec. 5 what role is played by the assumption that $\Sigma$ is in uniform motion relative to S. If the motion of $\Sigma$ (even if the tetrahedron $\sigma_0\sigma_1\sigma_2\sigma_3$ stays ``rigid'') were arbitrary, not much simple could be said about the relation between the times at $\sigma_0$ of the two systems, and hence, of the relations between the co-ordinates of any event by them. However, as seen in Sec. 6.3 above, if we have one more system $\Sigma'$ in uniform motion at velocity $\fb{w}$ relative to S, then it will be also in uniform motion relative to $\Sigma$ (provided $\fb{w} \cdot \fb{v} \ne 1$), so that in this sense we have the group property for the set of observers in uniform motion relative to one another. But the relative times and co-ordinates are determined only within an unknown multiplier like $\beta_1$. It is convenient, of course, to assume, with Einstein, that
\[
\beta_1 = \beta = \sqrt{1-v^2/c^2},
\]
but then only an omniscient observer could verify whether this is so or not. 

\subsection{Length ``Contraction'' and Time ``Dilatation''}
We have the relation 
\[
\begin{array}{lll}
\delta_1=\gamma \sqrt{(1-v^2)d_1^2+(\fb{d_1}\cdot \fb{v})^2},
\end{array}
\]
where $\gamma = \beta_1 / (1-v^2)$ and so, the distance between $\sigma_0$ and $\sigma_1$ in S, namely $d_1$, need not be the same as the distance between them in $\Sigma$, unless $\beta_1$ is chosen ``properly''. But then we have the other two relations also to worry about:
\[
\begin{array}{lll}
\delta_2&=&\gamma \sqrt{(1-v^2)d_2^2+(\fb{d_2} \cdot \fb{v})^2} \\
\delta_3&=&\gamma \sqrt{(1-v^2)d_3^2+(\fb{d_3} \cdot \fb{v})^2}
\end{array}
\]
and so we cannot have all the equalities $\delta_1=d_1$, $\delta_2=d_2$, $\delta_3=d_3$ unless $\fb{v}=\fb{0}$ and $\beta_1=1$. 

But then what is this contraction or change in ``length'' in our approach? As we have remarked above, distance in S and distance in $\Sigma$ are non-commensurate or independent \un{concepts}, though their numerical values could be related.

A similar comment could be made with regard to time dilatation. Time (clock) in S and time (clock) in $\Sigma$ are independent concepts, although, if we assume the ``rigidity'' condition, they are numerically related; thus, \un{at} $\sigma_0$
\[
\tau=\beta_1 t.
\]
But the coefficient $\beta_1$ is entirely arbitrary, or rather will be known only to an omniscient observer who can read both the clocks. If we choose $\beta_1$ so that $\phi(v)=\beta_1/\sqrt{1-v^2}=1$, with Einstein, then we do have $\tau=\sqrt{1-v^2}t$, a case of time dilatation. With this choice, it turns out that \un{at} $s_0$ in S, there is a time dilatation by the same factor, i.e., $t=\sqrt{1-v^2}\tau$.

\subsection{The ``Twin Paradox''}
We observe that neither in Einstein's approach nor in our approach can anything be said about what will happen when an observer, \un{previously} at rest, is \un{set} in motion. Although Einstein does say that ``now to the origin of one of the two systems (k) let a constant velocity v be imparted in the direction of the increasing $x$ of the other stationary system (K)'', his derivations \un{nowhere} use this conception. It is unfortunate that this way of putting it seems to have led him to formulate what has become known, after Langevin, as the ``twin paradox''. To emphasize again, Einstein's theory does not say anything as to what happens when a ``clock'' is set in motion. In our approach, we start with the premise that there are two observers, and that one of them is (already) in motion relative to the other. We, therefore, feel that the speculations by Einstein are not justified. Further, for the clock to return to its starting point, he had to ``assume that the result proved for a polygonal line is also valid for a continuously curved line''. 

\subsection{Invariance of other Laws of Physics}
It is surprising that Einstein was not tempted to consider some other Laws of Physics for an application of his ``Principle of Relativity''. Does the ``Principle of Relativity'' apply to another time-honored Law of Physics, much older than the Maxwell-Hertz Law, namely, \emph{Newton's Law of Universal Gravitation}? Or, an even earlier, simpler, Law, namely, \emph{Hooke's Law}? Both these laws involve the concept of \un{simultaneity at a distance} because both of them refer to the positions of two bodies at the same instant of time. (Levich \cite{LEVICH} says : ``\ldots the theory of relativity is incompatible with the notion of action at a distance. Two events can be in a mutual relationship as cause and effect only where they occur at the same place simultaneously as is required by the concept of short-range action.)  In the Law of Universal Gravitation, the instantaneous force on each of several mutually gravitating bodies depends (in the inverse square manner) on the distances of that body from the other bodies at that \un{same instant}. Similarly, for a massless spring, the force exerted by the spring at each end at each instant depends on the distance between the two ends at that instant. Now, recalling Einstein's observation ``that two events which, viewed from a system of co-ordinates, are simultaneous, can no longer be looked upon as simultaneous events when envisaged from a system which is in motion relatively to that system'', we see that the matter of invariance of Newton's Law of Universal Gravitation and of Hooke's Law (perhaps along with Newton's Third Law of Motion) requires investigation. 

We can show easily that Hooke's Law does not satisfy the ``Principle of Relativity'', that is to say, invariance of form under the Lorentz-Einstein transformation. 
First, we show that the \un{form} of variation of a physical variable may not remain invariant under the Lorentz-Einstein transformation. Suppose a particle has a sinusoidal motion in the x-direction of the stationary system K, its position $x(t)$ at time $t$ being given by 
\shrteqn{
x(t) = \sin(t)
}
(For example, such would be the classical frictionless motion of a point mass connected to a spring whose other end is fixed.) Considering the special case when system k moves relatively to K in the direction of the x-axis of K with velocity $v$, we have the familiar relations : 
\shrteqn{
\begin{array}{lll}
\tau &=& \beta (t - vx/c^2),\\
\xi &=& \beta (x - vt).
\end{array}
\label{eqn:local1}
}
What will be the form of motion in system k? Will it be sinusoidal also, i.e., given by 
\shrteqn{
\xi(\tau) = a \sin(\omega\tau+\phi)
\label{eqn:local2}
}
for some constants $a$, $\omega$ and $\phi$ ? The answer is : ``no'', as will be seen by substituting for $\tau$ and $\xi$, using \eqnrefs{eqn:local1}, in \eqnref{eqn:local2}, since we do not obtain an identity. (Interestingly, MacColl \cite{MACCOLL} shows that with the relativistic variation of mass, the motion of the mass-spring system is not sinusoidal.)

Similarly, a uniformly accelerated motion in K does not remain uniformly accelerated in k. However, a uniform motion in K remains uniform in k - as we know already from the Law of Composition of Velocities.

Perhaps, one could modify Laws such as Hooke's Law and Newton's Law of Universal Gravitation by using the pre-Einsteinian idea of ``retarded argument''.

\section{A NEW LOOK AT THE ``ELECTRODYNAMICAL PART'' OF EINSTEIN'S PAPER}
It seems that Einstein may have had (at least) the following different motivations in writing his \cite{PRI} (not necessarily in the order of their importance for \un{him}) : 
\begin{enumerate}
\item giving operational meaning to the ``time'' of a remote event, unlike the ``Ortzeit'' of Lorentz;
\item deriving the ``theory of transformation of co-ordinates and times'', using this operational meaning of time and the sameness of the velocity of light in the two systems, independent of the velocity of the emitting body;
\item deriving the invariance of the form of one particular ``law'' of physics, namely, the Maxwell-Hertz equations of the electromagnetic field (his ``Principle of Relativity''); and, of course,
\item deriving several new results.
\end{enumerate} 

Einstein, however, did not give an operational meaning to the \un{co-ordinates} of a remote event. We have shown how this could be done. As mentioned above, it seems it is not necessary to think of the second system of co-ordinates and time as being ``real''. It is enough to \un{model} the second system within the first system.

As we show in Sec. 7.2 below, the invariance of the form of the Maxwell-Hertz equations does \un{not} follow from the theory of transformation of co-ordinates and times. Rather, if we assume the invariance \un{and} Planck's formula for the transformation of \un{mass}, then the Lorentz force equation remains invariant! Thus, the surprising outcome is that insistence on the invariance of some laws (Maxwell-Hertz equations, Lorentz force equation) suggests a change in the \un{formulation} of some \un{other} law, namely, Newton's Second Law of Motion, or more specifically, in the expression for ``accelerative force'' in Newton's Second Law.

The first part of Einstein's paper, titled ``The Kinematical Part'', is really about the relation between time and space determination of events in two different observation systems. (In our treatment, the space co-ordinatization is a defined concept.) However, usually, it has been taken to be about ``transformation of co-ordinates''. This being the case, in electrodynamics, and in particular, as far as Maxwell's equations are concerned, what are the \un{events} being studied? The X, Y, Z components of the electric field and the L, M, N components of the magnetic field are not events! In fact, there could be a ``vicious cycle'' here since the very determination of time and space uses light (signal) which, following Maxwell, is believed to be an electromagnetic phenomenon. Even the application of the ``relativistic'' approach to mechanics will engender light in the observation of mechanical phenomena. But then this would need a new approach to electrodynamics and the ``winning'' of Maxwell's equations.

\subsection{Einstein's ``New Manner of Expression'' and ``Dynamics of the slowly accelerated Electron''}
Taking Maxwell's equations for granted, as does Einstein, it appears that there is a flaw in the section ``Transformation of the Maxwell-Hertz Equations for Empty Space.  On the Nature of the Electromotive Forces Occurring in a Magnetic Field During Motion'' in the ``Electrodynamical Part'' of Einstein's paper. To point it out, we need to quote him at length. 
\\
\\
``Let the Maxwell-Hertz equations for empty space hold good for the stationary system K, so that we have

\renewcommand{\arraystretch}{1.5}
{\Large
\[
\begin{array}{llllll}
\frac{1}{c}\dd{\rm X}{t} & = & \dd{\rm N}{y} - \dd{\rm M}{z}, &
    \frac{1}{c}\dd{\rm L}{t} & = & \dd{\rm Y}{z}-\dd{\rm Z}{y}, \\
\frac{1}{c}\dd{\rm Y}{t} & = & \dd{\rm L}{z} - \dd{\rm N}{x}, &
    \frac{1}{c}\dd{\rm M}{t} & = & \dd{\rm Z}{x}-\dd{\rm X}{z}, \\
\frac{1}{c}\dd{\rm Z}{t} & = & \dd{\rm M}{x} - \dd{\rm L}{y}, &
    \frac{1}{c}\dd{\rm N}{t} & = & \dd{\rm X}{y}-\dd{\rm Y}{x}, \\
\end{array}
\]
}
\renewcommand{\arraystretch}{1}

\noindent
where (X, Y, Z) denotes the vector of the electric force, and (L, M,
N) that of the magnetic force.

If we apply to these equations the transformation developed in \S\ 3, by
referring the electromagnetic processes to the system of co-ordinates
there introduced, moving with the velocity $v$, we obtain the equations

\renewcommand{\arraystretch}{1.5}
{\large
\[
\begin{array}{rcll}
\frac{1}{c}\dd{\rm X}{\tau} & = & \dd{}{\eta}\left\{\beta\left({\rm N}-\frac{v}{c}{\rm Y}\right)\right\} & -\dd{}{\zeta}\left\{\beta\left({\rm M}+\frac{v}{c}{\rm Z}\right)\right\}, \\
\frac{1}{c}\dd{}{\tau}\left\{\beta\left({\rm Y}-\frac{v}{c}{\rm N}\right)\right\} & = & \dd{\rm L}{\xi} & - \dd{}{\zeta}\left\{\beta\left({\rm N}-\frac{v}{c}{\rm Y}\right)\right\}, \\
\frac{1}{c}\dd{}{\tau}\left\{\beta\left({\rm Z}+\frac{v}{c}{\rm M}\right)\right\} & = & \dd{}{\xi}\left\{\beta\left({\rm M}+\frac{v}{c}{\rm Z}\right)\right\} & - \dd{\rm L}{\eta}, \\
\frac{1}{c}\dd{\rm L}{\tau} & = & \dd{}{\zeta}\left\{\beta\left({\rm Y}-\frac{v}{c}{\rm N}\right)\right\} & - \dd{}{\eta}\left\{\beta\left({\rm Z}+\frac{v}{c}{\rm M}\right)\right\}, \\
\frac{1}{c}\dd{}{\tau}\left\{\beta\left({\rm M}+\frac{v}{c}{\rm Z}\right)\right\} & = & \dd{}{\xi}\left\{\beta\left({\rm Z}+\frac{v}{c}{\rm M}\right)\right\} & -\dd{\rm X}{\zeta}, \\
\frac{1}{c}\dd{}{\tau}\left\{\beta\left({\rm N}-\frac{v}{c}{\rm Y}\right)\right\} & = & \dd{\rm X}{\eta} & - \dd{}{\xi}\left\{\beta\left({\rm Y}-\frac{v}{c}{\rm N}\right)\right\}, \\
\end{array}
\]
}
\renewcommand{\arraystretch}{1}

\noindent
where 

\[
\beta = 1/\sqrt{1-v^2/c^2}.
\]

Now the principle of relativity requires that if the
Maxwell-Hertz equations for empty space hold good in system K,
they also hold good in system $k$; that is to say that the
vectors of the electric and the magnetic force---(\pr{X},
\pr{Y}, \pr{Z}) and (\pr{L}, \pr{M}, \pr{N})---of the moving
system $k$, which are defined by their ponderomotive effects on
electric or magnetic masses respectively, satisfy the following
equations:---

\renewcommand{\arraystretch}{1.5}
{\Large
\[
\begin{array}{cccccc}
\ic\dd{\rm X'}{\tau} & = & \dd{\rm N'}{\eta}-\dd{\rm M'}{\zeta}, & \ic\dd{\rm L'}{\tau} & = & \dd{\rm Y'}{\zeta} - \dd{\rm Z'}{\eta}, \\
\ic\dd{\rm Y'}{\tau} & = & \dd{\rm L'}{\zeta}-\dd{\rm N'}{\xi}, & \ic\dd{\rm M'}{\tau} & = & \dd{\rm Z'}{\xi} - \dd{\rm X'}{\zeta}, \\
\ic\dd{\rm Z'}{\tau} & = & \dd{\rm M'}{\xi}-\dd{\rm L'}{\eta}, & \ic\dd{\rm N'}{\tau} & = & \dd{\rm X'}{\eta} - \dd{\rm Y'}{\xi}. \\
\end{array}
\]
}
\renewcommand{\arraystretch}{1}

Evidently the two systems of equations found for system $k$ must express
exactly the same thing, since both systems of equations are equivalent
to the Maxwell-Hertz equations for system K\@. \emph{Since, further, the
equations of the two systems agree, with the exception of the symbols
for the vectors, it follows that the functions occurring in the
systems of equations at corresponding places must agree, with the
exception of a factor $\psi(v)$, which is common for all functions of the
one system of equations, and is independent of $\xi, \eta, \zeta$ and $\tau$ but depends
upon $v$.}'' [our italics].
\\
\\

Is the last sentence of the quotation above (our italics) a valid conclusion from the preceding discussion? (Einstein's notation could cause some confusion; he uses the same letter, $\rm X$, for example, to denote both a function of $(x,\ y,\ z,\ t)$ and of $(\xi,\ \eta,\ \zeta,\ \tau)$, having, of course, the same value at the corresponding quadruples $(x,\ y,\ z,\ t)$ and $(\xi,\ \eta,\ \zeta,\ \tau)$.) Indeed, two pages later (``Theory of Doppler's Principle and of Aberration''), Einstein uses a non-zero solution of the Maxwell-Hertz equations in free space.  So, all that Einstein is entitled to say is that the \un{differences} $\rm X'-\rm X$, $\rm Y'-\beta(\rm Y-\frac{v}{c}\rm N)$, $\rm Z'-\beta(\rm Z+\frac{v}{c}\rm M)$, $\rm L'-\rm L$, $\rm M'-\beta(\rm M+\frac{v}{c}\rm Z)$, $\rm N'-\beta(\rm N-\frac{v}{c}\rm Y)$ must satisfy the Maxwell-Hertz equations.

Of course, although it is not \un{necessary} that $\rm X'-\rm X=0$, $\rm Y'-\beta(\rm Y-\frac{v}{c}\rm N)=0$, etc., it is \un{sufficient} in the sense if we \un{define} the new functions $\rm X'$, etc., by equations :

\[
\begin{array}{cclccl}
{\rm X'} & = & \psi(v){\rm X}, & {\rm L'} & = & \psi(v){\rm L}, \\
{\rm Y'} & = & \psi(v)\beta\left({\rm Y}-\frac{v}{c}{\rm N}\right), & {\rm M'} & = & \psi(v)\beta\left({\rm M}+\frac{v}{c}{\rm Z}\right), \\
{\rm Z'} & = & \psi(v)\beta\left({\rm Z}+\frac{v}{c}{\rm M}\right), & {\rm N'} & = & \psi(v)\beta\left({\rm N}-\frac{v}{c}{\rm Y}\right). \\
\end{array}
\]
or, accepting Einstein's argument that $\psi(v)=1$, by equations which can be legitimately called the ``Einstein Field Transformation Equations'': 
\shrteqn{
\begin{array}{cclccl}
{\rm X'} & = & {\rm X}, & {\rm L'} & = & {\rm L}, \\
{\rm Y'} & = & \beta\left({\rm Y}-\frac{v}{c}{\rm N}\right), & {\rm M'} & = & \beta\left({\rm M}+\frac{v}{c}{\rm Z}\right), \\
{\rm Z'} & = & \beta\left({\rm Z}+\frac{v}{c}{\rm M}\right), & {\rm N'} & = & \beta\left({\rm N}-\frac{v}{c}{\rm Y}\right). \\
\end{array}
}
then these new functions would describe a field in the moving system that would satisfy the Maxwell-Hertz equations as ``required'' by the Principle of Relativity. Thus, the Principle of Relativity is a \un{guiding principle} rather than a \un{physical law}. 

Einstein goes on to interpret the field transformation equations. We quote again:
\\
\\
``Consequently the first three equations above allow themselves to be
clothed in words in the two following ways:---

1.  If a unit electric point charge is in motion in an electromagnetic
field, there acts upon it, in addition to the electric force, an
``electromotive force'' which, if we neglect the terms multiplied by the
second and higher powers of $v/c$, is equal to the vector-product of the
velocity of the charge and the magnetic force, divided by the velocity
of light.  (Old manner of expression.)

2.  If a unit electric point charge is in motion in an electromagnetic
field, the force acting upon it is equal to the electric force which
is present at the locality of the charge, and which we ascertain by
transformation of the field to a system of co-ordinates at rest
relatively to the electrical charge.  (New manner of expression.)''
\\
\\
Now, his ``old manner of expression'' corresponds to the Lorentz force equations. What may have caused Einstein to think of the ``new manner of expression''? If $(x(t),\ y(t),\ z(t))$ denotes the instantaneous position of the charge in S, the $x$-, $y$- and $z$- components of the force produced by the field are then given by the Lorentz equations : 
\shrteqn{
\begin{array}{lll}
\displaystyle F_x &\displaystyle  = & \displaystyle \rm X + \frac{\dot{y}}{c}\rm N - \frac{\dot{z}}{c}\rm M\ ,\\[0.15in]
\displaystyle F_y &\displaystyle  = & \displaystyle \rm Y + \frac{\dot{z}}{c}\rm L - \frac{\dot{x}}{c}\rm N\ ,\\[0.15in]
\displaystyle F_z &\displaystyle  = &\displaystyle  \rm Z + \frac{\dot{x}}{c}\rm M - \frac{\dot{y}}{c}\rm L\ .
\label{eqn:lorentz}
\end{array}
}
If we put $\dot{x}=v$, $\dot{y}=0$, $\dot{z}=0$, then the expressions on the right-hand-side of the equations above look almost like the expressions on the right-hand-side of the field transformation equations. 

His ``new manner of expression'' has charmed a number of authors because it seems to \un{reduce} electrodynamics to electrostatics. But it does not seem to have been realized that the new manner is not useful when the electric charge does not have a uniform motion, or when there is more than one moving charge. Further, we show below that we can hold on to the Lorentz force equations for arbitrary motion of the charge in the stationary system because they will hold in the moving system too provided we make an important change, as suggested by Planck, in the way we handle ``mass''. Einstein himself seems to have been charmed by his new manner of expression so that he has to consider ``the slowly accelerated electron'' in the last section of his paper ``Dynamics of the Slowly Accelerated Electron'', and use language like :

``If the electron is at rest at a given epoch, the motion of the
electron ensues in the \emph{next instant of time} [our italics] according to the equations
\[
\begin{array}{lllllllllll}
m\frac{d^2x}{dt^2} & = & \epsilon{\rm X} &, & m\frac{d^2y}{dt^2} & = & \epsilon{\rm Y} & ,& m\frac{d^2z}{dt^2} & = & \epsilon{\rm Z} 
\end{array}
\]

\noindent
where $x, y, z$ denote the co-ordinates of the electron, and $m$ the mass
of the electron, as long as its motion is slow.

Now, secondly, let the velocity of the electron at a given epoch be $v$.
We seek the law of motion of the electron in the \emph{immediately ensuing
instants of time.}'' [our italics].

Einstein applies his field transformation theory to the motion of an electron, by noting that the field quantities $\rm X$, $\rm Y$, $\rm Z$, and $\rm X'$, $\rm Y'$, $\rm Z'$ do determine the force acting on the electron. But what is ``force'' acting on a moving body? Einstein finds out that if we ``maintain the equation---mass $\times$ acceleration $=$ force'', then the electron has two different masses : 
\begin{eqnarray*}
{\rm Longitudinal\ mass} & = & \frac{m}{(\sqrt{1-v^2/c^2})^3}\ , \\
{\rm Transverse\ mass} &  = & \frac{m}{1-v^2/c^2}\ .
\end{eqnarray*}

Of course, J.J.Thomson and others had deduced earlier that a moving electron has a velocity-dependent mass, but their approach was different from Einstein's. Einstein's approach hinges on his ``Theory of Transformation of Co-ordinates and Times'', which, in turn, follows from his two Postulates. Further, Einstein suggested: ``With a different definition of force and acceleration we should naturally obtain other values for the masses''. Perhaps, Planck\cite{PLANCK1} was inspired by this suggestion. Also, Einstein boldly asserted : ``\ldots these results as to the mass are also valid for ponderable material points, because a ponderable material point can be made into an electron (in our sense of the word) by the addition of an electric charge, \emph{no matter how small}.''

\subsection{Lorentz Force and ``variable'' Mass}
We show below that if \un{accelerative} force is defined as suggested by Planck\cite{PLANCK1}, then the equations of motion of a charge in the \un{stationary} system under the action of the Lorentz force imply the equations of motion of a charge in the \un{moving} system under the action of the Lorentz force. We are also able to see a ``reason'' why a dynamics in which the mass of a charged body is constant, independent of the velocity, is not compatible with the invariance of the Maxwell-Hertz equations and the Lorentz force equations. It should be pointed out that Planck, in his derivation, surprisingly says that $v$ is to be replaced by $\sqrt{\dot{x}^2 + \dot{y}^2 + \dot{z}^2}$ in \eqnrefs{eqn:Einstein} and \eqnrefs{eqn:LorentzForce} below (``indem \"uberall q an die Stelle von $v$ gesetzt wird'', Planck's ``q'' being $\sqrt{\dot{x}^2 + \dot{y}^2 + \dot{z}^2}$).

We will consider here only the special case studied by Einstein where the moving system $\Sigma$ (Einstein's ``$k$'') moves with a constant velocity $v$ in the direction of the $x$-axis of the stationary system S (Einstein's K). Let us assume that the field quantities X, Y, Z, L, M, N determine the force acting on a unit moving charge as given by \un{Lorentz's formula}. If $(x(t),\ y(t),\ z(t))$ denotes the instantaneous position of the charge in S, the $x$-, $y$- and $z$- components of the force produced by the field are then given by \eqnrefs{eqn:lorentz}.

The motion of the charge as seen by the moving observer $\Sigma$ is given by $(\xi(\tau),\ \eta(\tau),\ \zeta(\tau))$ where 
\shrteqn{
\begin{array}{lll}
\tau & = &\beta(t-vx(t)/c^2),\\
\xi(\tau)&=&\beta(x(t)-vt),\\
\eta(\tau)&=&y(t),\\
\zeta(\tau)&=&z(t),
\label{eqn:Einstein}
\end{array}
}
where $\beta=1/\sqrt{1-v^2/c^2}$. The velocity components in $\Sigma$ turn out to be\cite{ROSSER} : 
\shrteqn{
\begin{array}{llllllllllllllllll}
\displaystyle
\frac{d\xi}{d\tau} & \displaystyle = & \displaystyle \dot{\xi} &\displaystyle = &\displaystyle \frac{\dot{x}-v}{u}&,&
\displaystyle
\frac{d\eta}{d\tau} &\displaystyle = &\displaystyle \dot{\eta} &\displaystyle = \displaystyle &\displaystyle \frac{\dot{y}}{\beta u}&,&
\displaystyle
\frac{d\zeta}{d\tau} &\displaystyle = &\displaystyle \dot{\zeta} &\displaystyle = &\displaystyle \frac{\dot{z}}{\beta u}&,
\end{array}
}
where $u=1-(v/c^2)\dot{x}$, and the accelerative components are given by
\shrteqn{
\begin{array}{llllllllllll}
\displaystyle \ddot{\xi} & \displaystyle= & \displaystyle\frac{\ddot{x}}{\beta^3 u^3}&,&
\displaystyle\ddot{\eta} &\displaystyle = & \displaystyle\frac{u\ddot{y}+\frac{v}{c^2}\dot{y}\ddot{x}}{\beta^2 u^3}&,&
\displaystyle\ddot{\zeta} & \displaystyle= &\displaystyle \frac{u\ddot{z}+\frac{v}{c^2}\dot{z}\ddot{x}}{\beta^2 u^3}&.
\end{array}
}
(In his study of the slowly accelerated electron, Einstein effectively sets ``at a given epoch'' $\dot{x}=v$, $\dot{y}=0$, $\dot{z}=0$ but $\ddot{x}$, $\ddot{y}$, $\ddot{z}$ may not be zero at that epoch, so that $u=\frac{1}{\beta^2}$, $\ddot{\xi}=\beta^3 \ddot{x}$, $\ddot{\eta}=\beta^2 \ddot{y}$, $\ddot{\zeta}=\beta^2 \ddot{z}$.)

If we now choose for $\Sigma$ the Lorentz forces given by 
\begin{eqnarray}
F_{\xi} & = & \rm X' + \frac{\dot{\eta}}{c}\rm N' - \frac{\dot{\zeta}}{c}\rm M',\\[0.10in]
F_{\eta} & = & \rm Y' + \frac{\dot{\zeta}}{c}\rm L' - \frac{\dot{\xi}}{c}\rm N',\\[0.10in]
F_{\zeta} & = & \rm Z' + \frac{\dot{\xi}}{c}\rm M' - \frac{\dot{\eta}}{c}\rm L',
\end{eqnarray}
where $\rm X'$, $\rm Y'$, $\rm Z'$, $\rm L'$, $\rm M'$, $\rm N'$ as functions of $(\xi,\ \eta,\ \zeta,\ \tau)$ are related to X, Y, Z, L, M, N by the Einstein relations, then we find (with Planck)\cite{PLANCK2} that these are related to the Lorentz forces in S as follows:
\shrteqn{
\begin{array}{lll}
\displaystyle {F}_{\xi} &\displaystyle  = &  \displaystyle {F}_x - \frac{v}{u c^2}\dot{y}{F}_y - \frac{v}{u c^2}\dot{z}{F}_z , \\
\displaystyle {F}_{\eta} & \displaystyle = & \displaystyle \frac{1}{\beta u} {F}_y, \\
\displaystyle {F}_{\zeta} &\displaystyle  = & \displaystyle \frac{1}{\beta u} {F}_z.
\label{eqn:LorentzForce}
\end{array}
}
Note that these relations \un{do not involve mass}. (These relations were derived by Planck\cite{PLANCK2} exactly as we have done above, and not on the basis of some other principles, as stated by Miller\cite{MILLER}.)
 
We now see that if we assume the mass of a charged body to be a constant, say, $m$, independent of its velocity, so that the accelerative force components are given by the product $(\rm {mass} \times \rm{acceleration\ \ component})$, then the equations of motion in S:
\shrteqn{
\begin{array}{llllllllllll}
m\ddot{x}&=&F_x&,&m\ddot{y}&=&F_y&,&m\ddot{z}&=&F_z&,
\end{array}
}
will not imply the equations of motion in $\Sigma$: 
\shrteqn{
\begin{array}{llllllllllll}
m\ddot{\xi}&=&F_{\xi}&,&m\ddot{\eta}&=&F_{\eta}&,&m\ddot{\zeta}&=&F_{\zeta}&.
\end{array}
}

Suppose that following Planck's suggestion we define the accelerative force to be the time-rate of change of momentum, assuming that the mass has a dependence on velocity given by
\shrteqn{
m(t) = {m_0}/\left (1-\frac{\dot{x}^2+\dot{y}^2+\dot{z}^2}{c^2}\right )^{1/2}
}
where $m_0$ is a constant, so that the accelerative force components in S are: 
\shrteqn{
\begin{array}{llllllllllll}
\overline{F}_x & = & \frac{d}{dt}(m\dot{x})&,&
\overline{F}_y & = & \frac{d}{dt}(m\dot{y})&,&
\overline{F}_z & = & \frac{d}{dt}(m\dot{z})&.
\end{array}
}

Assuming that the variation of mass in $\Sigma$ is given by  
\[
\overline{m}(\tau) = {m_0'}/{\left (1-\frac{\dot{\xi}^2+\dot{\eta}^2+\dot{\zeta}^2}{c^2}\right )^{1/2}}\ ,
\]
 the accelerative force components in $\Sigma$ are defined by 
\shrteqn{
\begin{array}{llllllllllll}
\overline{F}_{\xi} & = & \frac{d}{d\tau}(\overline m\dot{\xi})&,&
\overline{F}_{\eta} & = & \frac{d}{d\tau}(\overline m\dot{\eta})&,&
\overline{F}_{\zeta} & = & \frac{d}{d\tau}(\overline m\dot{\zeta})&.
\end{array}
}

Fortunately, on using the relations between the velocity and acceleration components in S with those in $\Sigma$ we find that with this definition of accelerative force, the accelerative forces in $\Sigma$ are related to the accelerative forces in S by relations, which are analogous to the relations between the Lorentz forces, as follows:
\shrteqn{
\begin{array}{lll}
\displaystyle \overline{F}_{\xi} & \displaystyle = & \displaystyle  \overline{F}_x - \frac{v}{u c^2}\dot{y}\overline{F}_y - \frac{v}{u c^2}\dot{z}\overline{F}_z\ , \\
\displaystyle \overline{F}_{\eta} &\displaystyle  = &\displaystyle  \frac{1}{\beta u} \overline{F}_y\ , \\
\displaystyle \overline{F}_{\zeta} &\displaystyle  = & \displaystyle \frac{1}{\beta u} \overline{F}_z\ ,
\label{eqn:PlanckRelations}
\end{array}
}
if and only if $m_0'=m_0$ (this corresponds to Einstein's conclusion that $\psi(v)=1$). (We have not seen  \eqnrefs{eqn:PlanckRelations} stated \un{explicitly} in the literature.) 

So, it follows immediately that the equations of motion in S: 
\shrteqn{
\begin{array}{llllllllllll}
\overline{F}_x &=& F_x&,&\overline{F}_y = F_y&,& \overline{F}_z = F_z&,
\end{array}
}
imply the equations of motion in $\Sigma$:
\shrteqn{
\begin{array}{llllllllllll}
\overline{F}_{\xi} &=& F_{\xi}&,&\overline{F}_{\eta} = F_{\eta}&,& \overline{F}_{\zeta} = F_{\zeta}&.
\end{array}
}

In fact, what we need is that the ratio $\displaystyle (\rm {mass\ constant}/\rm {charge})$ has the same value in S and $\Sigma$. If we assume that the charge has the same value in S and $\Sigma$, then the constant in the definition of accelerative force has to be the same for both S and $\Sigma$, i.e., to be independent of the observer. It could be termed the ``absolute mass'' or even the ``rest mass'', since it is the mass when the velocity is zero no matter in which system. (It may be better not to refer to quantities like $m(t)$ above as ``mass'' or ``variable mass''. What matters is how a co-ordinate dependent quantity ``accelerative force'' is defined in relation to a co-ordinate independent constant called (mass) and co-ordinate dependent position and time.)

\subsection{Maxwell-Hertz, Lorentz, Einstein, and Planck}
Thus, the Einstein transformation of co-ordinates and times, the Einstein transformation of field quantities, the Maxwell-Hertz equations, the Lorentz force equations, and the Planck definition of accelerative force all hang together as well in the stationary system S as in the moving system $\Sigma$. Einstein's ``new manner of expression'' for the force on a moving charge is not required at all and we do not have to agree with Einstein that the ``electromotive force plays in the developed theory merely the part of an auxiliary concept''.  ``Newtonian mechanics'' can be seen to be valid for charged bodies if instead of Newton's definition of accelerative force, we use Planck's definition. It is interesting that Einstein's insistence that ``the laws by which the states of physical systems undergo change are not affected, whether these changes of state be referred to the one or the other of two systems of co-ordinates in uniform translatory motion'' has led to a new definition of accelerative force. 

\subsection{Other Definitions of accelerative Force}
We might ask: are there other definitions of accelerative force which will work? We show immediately that if we assume that the mass is a differentiable function only of $(\dot{x}^2 + \dot{y}^2 + \dot{z}^2)$, then the only functions that will work are the Planck function and a function that we give below. Denoting $\sqrt{(\dot{x}^2 + \dot{y}^2 + \dot{z}^2)}$ by $q$ as before and a desired accelerative force function by $m(q)$, on substituting in \eqnrefs{eqn:PlanckRelations} and using the fact that $\dot{x}$, $\dot{y}$, $\dot{z}$ are arbitrary, in particular, $\dot{x}=v$, we obtain the following differential equation for $m$:
\[
\frac{dm}{dq}=\frac{mq}{(c^2-q^2)}
\]
the solution of which is, not $k_0 \{c^2-q^2\}^{-1/2} $, but 
\[
k_0 |c^2-q^2|^{-1/2} 
\]
except for $q=c$. Thus, the solution has two branches:
\[
m(q)=k_0 (c^2-q^2)^{-1/2}
\]
for $q<c$, and 
\[
m(q)=k_0 (q^2-c^2)^{-1/2}
\]
for $q>c$.

Thus, although the Planck formula for mass presumes that 
\[
\dot{x}^2 + \dot{y}^2 + \dot{z}^2 < c^2,
\]
i.e., that the body moves more slowly than light, if the body moves faster than light, i.e., if 
\[
\dot{x}^2 + \dot{y}^2 + \dot{z}^2 > c^2,
\]
we could use the formula : 
\shrteqn{
m(t) = {m_0}/{\left (\frac{\dot{x}^2+\dot{y}^2+\dot{z}^2}{c^2} - 1 \right )^{1/2}},
}
so that there is no need to agree with Einstein that ``Velocities greater than that of light have - as in our previous results - no possibilities of existence'' and to entertain any idea of ``purely imaginary'' mass for faster-than-light bodies. Incidentally, we find that 
\shrteqn{
1-\left (\frac{\dot{\xi}^2+\dot{\eta}^2+\dot{\zeta}^2}{c^2} \right ) 
=
\frac{1}{\beta ^2 u^ 2} \left [ 1 - \frac{\dot{x}^2+\dot{y}^2+\dot{z}^2}{c^2} \right ],
}
so that a body moves faster-than-light in S if and only if it moves faster-than-light in $\Sigma$. Perhaps this result can be proved in the general framework of Sec. 5.

\subsection{Charged Bodies Traveling at the Speed of Light}
Of course, there is a singularity in the formula for mass when the body moves as fast as light.  What should be the mass formula for the case $ \dot{x}^2 + \dot{y}^2 + \dot{z}^2 = c^2$? (It is possible that this equality may hold over an interval of time, and not at just one time instant.) Note that when $q=c$ we also have $\sqrt{\dot{\xi}^2+\dot{\eta}^2+\dot{\zeta}^2}=c$, so that the mass of such a charged body must be the same in both S and $\Sigma$, say, $m_c$. Assuming that the Newtonian definition of accelerative force holds in this case, we find that  \eqnrefs{eqn:PlanckRelations} will be satisfied only with $m_c=0$, i.e., the charged body has zero mass in S and $\Sigma$. In that case, the field must be such that the Lorentz force components are zero, i.e., in  \eqnrefs{eqn:lorentz}, $F_x=F_y=F_z=0$. The resulting system of equations for  $\dot{x}$, $\dot{y}$, $\dot{z}$ in terms of X, Y, Z, L, M, N has zero determinant. A simple calculation shows that for a solution to exist, the following condition must be satisfied :
\shrteqn{
\rm{XL} + \rm{YM} + \rm{ZN} =0\ ,
}
i.e., the electric field vector must be perpendicular to the magnetic field vector. Under this condition, we can solve the equations for $\dot{y}$, $\dot{z}$ in terms of $\dot{x}$ and the field components, and substitute these in the velocity condition: 
\shrteqn{
 \dot{x}^2 + \dot{y}^2 + \dot{z}^2 = c^2\ ,
\label{eqn:VelCondition}
}
obtaining a quadratic equation for $\dot{x}$: 
\shrteqn{
(\mathrm{L}^2 + \mathrm{M}^2 + \mathrm{N}^2 )\ \dot{x}^2 + 2c(\mathrm{MZ}-\mathrm{NY})\ \dot{x} + c^2(\mathrm{Y}^2 + \mathrm{Z}^2 - \mathrm{L}^2) = 0 \ .
\label{eqn:Quadrxdot}
}
This equation has real solutions if and only if 
\shrteqn{
\rm L^2 + \rm M^2 + \rm N^2 \ge \rm X^2 + \rm Y^2 + \rm Z^2\ ,
}
i.e., the magnetic field is at least as strong as the electric field. 

Thus, if the electric and magnetic fields are mutually perpendicular and the magnetic field is at least as strong as the electric field, the question of the possibility of the motion of zero-mass charged bodies at the speed of light amounts to the existence of the solution of  \eqnrefs{eqn:lorentz} with their left-hand-sides set equal to zero  alongwith the velocity condition \eqnref{eqn:VelCondition}. Note that the field components are to be evaluated \un{along} the motion. 

One can immediately verify that in the case of the field corresponding to a plane wave, a motion in straight line in any direction at the speed of light is possible in such a field. Could one conjecture that a photon may be a zero-mass charged body with an ``infinitesimally small'' charge, just as Einstein thought that ``a ponderable material point can be made into an electron (in our sense of the word) by the addition of an electric charge, \emph{no matter how small}''? (With zero accelerative force, the magnitude of the charge has no effect on the motion !) Thus, light, instead of \un{being} an electromagnetic wave, could consist of zero-mass charged particles moving \un{in} a suitable electromagnetic field. Could some appropriate field allow motion at the speed of light in a circle?

Finally, could there be a ``light barrier'' so that a motion cannot reach the speed of light even for one instant, whether from ``below'' or from ``above''?

\section{CONCLUDING REMARKS}
We have extended Einstein's admonition - ``a mathematical description of this kind has no physical meaning unless we are quite clear as to what we understand by ``time'' ''- to apply to the concept of ``place'' or ``co-ordinates'', i.e., of a co-ordinate system. We have shown how by considering a system S of an observer with a clock, aided by three reflecting stations, co-ordinates of a remote event can be \un{defined} and \un{determined}, as also the time of its occurrence. 

Considering another system $\Sigma$ ``in uniform motion of translation relatively to'' S, we have proved that the co-ordinates and times in $\Sigma$ are linearly related to the co-ordinates and times in S. The ``Lorentz transformation'' relating the two could be calculated and turns out to be identical with Einstein's formulas in the special case considered by him.

We have emphasized that the co-ordinates are a matter of \un{representation} of the observed data of \un{times} of transmission and reception of various signals by the observer. The representation we have used, and the one Einstein implicitly assumed, namely, 3-dimensional Euclidean geometry, is based on our ``experience of space'', but it is only a representation of the data. This is not to say, of course, that our experiences of \un{seeing} a remote object at a certain time in our clock at a certain place are not ``real''. Indeed, there should be no hesitation in saying that ``That train arrives here at 7 o'clock '', or that, ``I saw the occultation of Venus by the moon beginning at 7 p.m.''. 

Is it possible that we may have experiences of departure and arrival times of signals which cannot be represented in 3-dimensional geometry? Could we use some other representation even if the 3-dimensional geometric representation is possible? (We do not mean here alternative co-ordinate systems, such as the spherical-polar, etc.) Also, what we have represented are only certain ``points'' in the ``motion'' of light signals ; thus we have supposed that light ``leaves'' at a certain place at a certain time and ``arrives'' at another place at another time. Perhaps, we could try to represent or model the entire \un{path} of the light signal. This might lead to a different approach to the ``General Theory of Relativity''. 

We have pointed out what appears to be a flaw in the ``Electrodynamical Part'' of Einstein's paper. We have shown that we can use the Lorentz force formula in the stationary system as well as in the moving system, and that Einstein's interpretation of ``electromotive force'' as an auxiliary concept is not necessary. Further, faster-than-light motions can be considered without any difficulty.

Lastly, we should perhaps recall Einstein's admonition again with regard to the atomic domain, such as that of an electron, and seek for ourselves operational meanings of time and distance on the atomic scale. Would the same signal suffice for this purpose? What would be a ``clock''? What would be ``observed'' and what would be ``defined'' and ``determined'' in terms of what is observed?

\clearpage
\myfiga
\clearpage
\myfigb
\clearpage
\myfigc
\clearpage
\myfigd
\clearpage
\noindent
Figure 1: Signal from $\sigma_0$ to $\sigma_1$ and back.
\\
\\
Figure 2: Direct and indirect echo from P in S.
\\
\\
Figure 3: Direct and indirect echo from P in $\Sigma$.
\\
\\
Figure 4: Delays in a round trip in S.

\end{document}